\documentclass[twocolumn]{aastex61}
\usepackage{amsmath}
\usepackage{amssymb}

\accepted{18/09/2017}
\submitjournal{AJ}

\shorttitle{Near-UV OH Prompt Emission in the Innermost Coma of 103P/Hartley 2}
\shortauthors{La Forgia et al.}

\begin{document}

\title{Near-UV OH Prompt Emission in the Innermost Coma of 103P/Hartley 2}

\correspondingauthor{Fiorangela La Forgia}
\email{fiorangela.laforgia@unipd.it}

\author{Fiorangela La Forgia}
 \affil{Department of Physics and Astronomy, University of Padova, Vicolo dell'Osservatorio 3, 35122 Padova, Italy}

\author{Dennis Bodewits}
 \affil{Department of Astronomy, University of Maryland, College Park, MD 20742-2421, USA}
  
\author{Michael F. A'Hearn$\dagger$}
 \affil{Department of Astronomy, University of Maryland,
 College Park, MD 20742-2421, USA}
 \email{$\dagger$ Mike A'Hearn died on May 29, 2017}

\author{Silvia Protopapa}
 \affil{Department of Astronomy, University of Maryland,
 College Park, MD 20742-2421, USA}

 \author{Michael S. P. Kelley} 
 \affil{Department of Astronomy, University of Maryland,
 College Park, MD 20742-2421, USA}

 \author{Jessica Sunshine}
 \affil{Department of Astronomy, University of Maryland,
 College Park, MD 20742-2421, USA}

\author{Lori Feaga}
\affil{Department of Astronomy, University of Maryland,
 College Park, MD 20742-2421, USA}

 \author{Tony Farnham}
 \affil{Department of Astronomy, University of Maryland,
 College Park, MD 20742-2421, USA}

\begin{abstract}

\noindent The Deep Impact spacecraft fly-by of comet 103P/Hartley 2 
 occurred on 2010 November 4, one week after perihelion with a closest approach (CA) distance
 of about 700 km. We used narrowband images obtained by the Medium Resolution Imager (MRI) onboard 
 the spacecraft to study the gas and dust in the innermost coma. 
 We derived an overall dust reddening of 15\%/100 nm 
 between 345 and 749 nm and identified a blue enhancement in the dust coma 
 in the sunward direction within 5 km from the nucleus, 
 which we interpret as a localized enrichment in water ice. 
 OH column density maps show an anti-sunward enhancement throughout the encounter except for the 
 highest resolution images, acquired at CA, where a radial jet becomes visible in the innermost coma, 
 extending up to 12 km from the nucleus.
 The OH distribution in the inner coma is very different from that expected for a fragment species. 
 Instead, it correlates well with the water vapor map derived 
 by the HRI-IR instrument onboard Deep Impact \citep{AHearn2011}. 
 Radial profiles of the OH column density 
 and derived water production rates show an excess of OH emission during 
 CA that cannot be explained with pure fluorescence. 
 We attribute this excess to a prompt emission process where photodissociation 
 of H$_2$O directly produces excited OH*($A^2\it{\Sigma}^+$) radicals. 
 Our observations provide the first direct imaging of Near-UV 
 prompt emission of OH. 
  We therefore suggest the use of a dedicated filter centered at 318.8 nm to directly trace the water in the coma of comets.
 
\end{abstract}


\section{Introduction}
\vspace{0.3cm}

    \setcitestyle{citesep={;}}
    
\noindent The hyperactive Jupiter family comet 103P/Hartley 2 was the second target of  
NASA's Deep Impact (DI) spacecraft. 
On 2010 November 4 at 13:59:47 UTC DI passed this small comet 
at a distance of 694 km from the nucleus with a speed of 12.3 km s$^{-1}$ \citep{AHearn2011}. 
Hartley 2 was then at 1.064 au from the Sun and had passed its perihelion one week before.
Visible observations of the comet acquired by the spacecraft at closest approach (CA)
show the resolved nucleus, 
with collimated jets from active areas on the surface \citep{Thomas2013}. 
Spectral observations of the ambient coma show that H$_2$O gas is enhanced above the central waist,
while water ice and dust are spatially correlated with CO$_2$ jets  above the smaller lobe
\citep{AHearn2011,Protopapa2014}.

DI observations allowed us to 
investigate the very innermost regions of the gas coma where the first chemical processes transforming
the original parent molecules into fragment species take place.
Water is one of the main volatiles sublimating from the nucleus and the most
abundant molecule present in the inner gas coma.
There, H$_2$O molecules are destroyed primarily through photodissociation 
into OH radicals. 
OH produced in the ground state is consequently excited by solar 
radiation and decays through the $A^2\it{\Sigma}^+-X^2\it{\Pi}$ transition, 
producing a strong near-UV (NUV) resonance 
fluorescence emission (RFE) band at 308.5~nm 
\citep{Schleicher_AHearn1982,Schleicher_AHearn1988}.
Owing to the direct connection between H$_2$O and OH, 
this cometary NUV emission has been used for decades 
as a tracer of the production and distribution of water \citep{AHearn1995}.

The Medium Resolution Imager (MRI) \citep{Hampton2005} 
onboard DI spacecraft was equipped with 
a NUV OH filter which allows for the study of OH emission.
However, the data from Hartley 2 presented in this paper reveal
that some additional emission 
mechanisms are needed beyond the RFE for interpretation.
This will be further discussed in Sec. \ref{disc}.

The photolysis of water also produces OH*($A^2\it{\Sigma}^+$) in the first
electronically excited state
with strongly populated high vibrationally and rotationally excited 
levels \citep{Carrington1964, Harich2000}.
This reaction channel, called OH prompt emission (PE), has a smaller 
but non-negligible branching ratio of 3.6\%, 
compared to 78.4\% of OH that is produced into the ground state 
(for quiet Sun conditions;
\citep[and references therein]{Combi2004}). 
The excited OH* fragments have a short lifetime of about 10$^{-6}$ s 
\citep{BeckerHaaks1973} and also decay to the ground state
through the same $A^2\it{\Sigma}^+-X^2\it{\Pi}$ transition but with different rotational 
structure. Due to the different rotational transition, 
the PE appears broader in wavelength (309--312 nm) than the RFE.

Both the spatial distribution of OH and the rovibrational spectrum 
can be used to distinguish prompt emission and resonant fluorescence emission.

First, due to their very short radiative lifetime, the OH* radicals do not 
travel large radial distances with respect to the 
photodissociating parent molecule, and emit within 0.1 cm from the water 
molecule they originated from. 
For comparison the OH RFE lifetime, i.e. the inverse of the fluorescence 
efficiency, at 1 au varies between 780 and 4500 s, depending on the 
heliocentric velocity of the comet \citep{Schleicher_AHearn1988}.
The morphology of OH PE thus resembles the
water distribution rather than the more extended distribution of OH RFE, 
and PE may exceed RFE in the inner tens of kilometers around the nucleus
\citep{Bertaux1986, Budzien_Feldman1991, Bonev2004, Bonev2006}.

Second, laboratory experiments \citep{Carrington1964,BeckerHaaks1973} 
and models \citep{Budzien_Feldman1991} indicate that the OH PE spectrum 
differs from the OH RFE spectrum because different rotational 
levels are populated.

Thus, very high spatial or spectral resolution is necessary 
to separate the two emission mechanisms.

The NUV transition is usually followed by additional rotational-vibrational 
transitions decaying toward lower rotational levels that contribute to
an IR band at $\sim$ 3 $\mu$m.
OH PE has been measured in both IR and NUV regions but never imaged directly. 
The 3 $\mu$m PE band has been detected in several ground-based observations 
of comets \citep[see for example][]{Brooke1996,Mumma2001,Gibb2003,Bonev2004,Bonev2006}. 
The NUV prompt emission at $308.5$ nm was originally studied 
by \citet{Bertaux1986} who suggested the presence of a bright spot 
of $\sim$ 33 km in the inner coma of comet C/1983 H1 (IRAF-Araki-Alcock), 
and evidence of the PE in the same comet was detected in observations 
by the International Ultraviolet Explorer \citep{Budzien_Feldman1991}. 
Recently \citet{AHearn2015} reported the first spectrally 
resolved detection of NUV OH prompt emission in comet C/1996 B2 (Hyakutake). 

In this paper we 
used narrowband filter observations obtained by MRI instrument
on board DI to study the spatial distribution of OH
in the innermost coma of Hartley 2, and we investigate 
the possibility that OH prompt emission is responsible for the observational 
evidences.

\newpage
\section{Observations and Data Processing}

\setcitestyle{citesep={;}}
\bibpunct[ ]{(}{)}{,}{a}{}{}

The MRI camera is a 2.1 m focal length Cassegrain telescope, with a 12 cm aperture, 
a field of view of approximately 35 $\times$ 35 
arcminutes, and a per pixel resolution of 2~arcsec \citep[see][for details]{Hampton2005}.
It is equipped with a total of nine filters, 
five of which are based on the Hale-Bopp narrowband filter set \citep{Farnham2000}.
Three narrowband filters are designed to measure different gas species 
(OH at 309.48~nm and CN at 388.80~nm with bandwidths 6.2 nm, 
and C$_2$ at 515.31~nm with bandwidth 11.8 nm), 
and two other narrowband filters are designed to measure the continuum colors
at 345 nm (Violet) and 526 nm (Green); two 
mediumband filters are designed to measure colors at 750 nm (Red) and 950 nm (IR).
The instrument is also equipped with two nearly identical broadband filters 
(Clear1 and Clear6) sensitive to the whole 200 - 1100 nm wavelength range
\citep{Klaasen2008}. 
The bandpass of the Clear filters includes continuum from light reflected by dust in the coma, as well 
as emission features from several fragment species, such as OH, CN, and C$_2$.  
Since narrowband filters require a relatively long exposure time to get a good signal to 
noise, during most of the encounter the Clear1 filter was used in
order to get an optimal sampling of the comet's lightcurve. 
 OH, CN and C$_2$ observations and some color observations
 were acquired from the day of the perihelion, i.e. 2010 October 28
 (DOY 301), through CA occurred on 2010 November 4 (DOY 308), and until 
2010 November 16 (DOY 320).

We analyzed a total of 153 OH images ranging from 
2010 October 28 to 2010 November 7 (DOY 301-311), 
acquired from a distance ranging between 10$^6$ 
and 8300 km, with fields of view 
(FOV) between 10$^4$ and 83 km (spatial 
scales at the comet
between 10$^4$ and $\sim$ 83 m/pixel).
We then focused on the two highest resolution images: 
image ID 5002027 obtained on DOY 308 about 11 minutes before encounter (E-11 min)
from a distance of 8303 km 
and image ID 5006065 acquired 8.5 minutes after encounter
(E+8.5 min) from 8256 km, 
having both a FOV of about 83 km and a spatial scale of 83 m/pixel.
Color images in the Violet, Green and Red filters were acquired close in time
to the OH observations (see Table \ref{tab.images}), allowing us to study the colors of the continuum and 
thus perform an accurate subtraction of the continuum light passing
through OH bandpass filter.

\setcitestyle{yysep={,}}
    \setcitestyle{citesep={;}}

  The MRI dataset used \citep{McLaughlin2011}  
  was calibrated using the pipeline described in \citet{Klaasen2008, Klaasen2013}.
 
As extensively discussed in \citet{Klaasen2013}, original images showed 
residual stripes caused by electrical interference.
They implemented an accurate de-striping pipeline which was applied to the data. 
Despite that, the residual gas data 
still shows some stripes, probably because of the very low signal to noise.
The stripes have a four quadrants pattern in full frame images, due to the 4 amplifiers used to read the CCD.
We implemented a further de-striping algorithm in addition to the one implemented
in the standard pipeline which takes the average of the first 50 rows from the edge
of each quadrant, then fits the stripe pattern with a polynomial function, 
and finally removes the pattern from the whole quadrant.

Coma structures are usually very faint and hidden by the isotropic component 
distribution. 
There are several enhancement techniques that can be used to highlight 
the anisotropies \citep[see for example][]{Samarasinha2011,Farnham2009}.
Here, we relied on one of the widely used methods to improve the contrast 
of the come morphology: the subtraction of the azimuthally averaged 
radial profile.
By subtracting this profile from the original image, 
anisotropic features of the coma that are otherwise concealed under a 
brighter isotropic coma,
are revealed and a more detailed analysis of the structures and of the 
spatial distribution is possible \citep[see for example][]{Schleicher2004}.
To extract radial profiles, 
images are converted into polar coordinates ($\rho$,$\Theta$)
where $\rho$ is the distance in km from the nucleus center, and $\Theta$ is the 
azimuth angle measured counterclockwise from the vertical line going from the center
of the nucleus perpendicularly to the bottom of the image.
 The profile is fitted with a polynomial function, which is converted 
into an image and subtracted from the gas image. 

 \section{Continuum Profiles and Colors}
 
     \setcitestyle{citesep={;}}
     
The coma brightness in the narrowband gas filters comes from 
two main components: gas emission lines, 
and continuum light scattered by the solid materials 
in the coma such as dust and ice. 
To isolate the gas emission contribution  
we need to estimate that of the dust.
In visible wavelengths, cometary dust is generally characterized by a 
featureless spectrum,
with a red slope with respect to the solar spectrum 
\citep[see for example][]{Jewitt1986}. 
We used the Violet (345 nm), Green (526 nm) and Red (750 nm)
color images acquired together with OH images (see Table \ref{tab.images})
to analyze the colors of the coma and
subtract the continuum contribution from 
OH images. 

\begin{deluxetable}{llccc}
\tablenum{1}
\tablecaption{OH images acquired at 14:48 UT (E-11 min)
 and at 14:10 UT (E+8.5 min) on 2010 Nov 4 and color images associated.\label{tab.images}}
\tablewidth{0pt}
\tablehead{
\colhead{image ID} & \colhead{UT} & \colhead{filter} & \colhead{range} & \colhead{FOV} \\
\colhead{} & \colhead{} & \nocolhead{} & \colhead{[km]} & \colhead{[km]}
}
\startdata
  5002027 & 13:48:25 & OH & 8303 & 83\\
  5002029 & 13:48:55 & VIOLET & 7883 & 79 \\
  5002034 & 13:50:06 & RED & 7185 & 72  \\
  5002035 & 13:50:10 & GREEN & 7102 & 71 \\
\hline
 5006060 & 14:09:07 & GREEN & 6982 & 70  \\
 5006061 & 14:09:18 & RED & 7071 & 71 \\
 5006064 & 14:09:54 & VIOLET & 7699 & 77 \\
 5006065 & 14:10:27 & OH & 8257 & 83 \\
\enddata
 \end{deluxetable}

These images are very close in time to the OH images (less than 2 minutes)
and despite the spacecraft movement during CA, the viewing geometry 
did not change significantly among them.

We rescaled and aligned colors images to the OH image, 
down to the sub-pixel level. 
Since the nucleus is resolved for these images, the alignment is done
using a projection of a shape model \citep{Thomas2013} 
of the nucleus in the image plane.
In order to avoid inaccuracies in the pointing,
the image alignment is further improved using an iterative cross correlation
process at the precision of a quarter of a pixel.

\begin{figure}[h!]
\epsscale{1.2}
\plotone{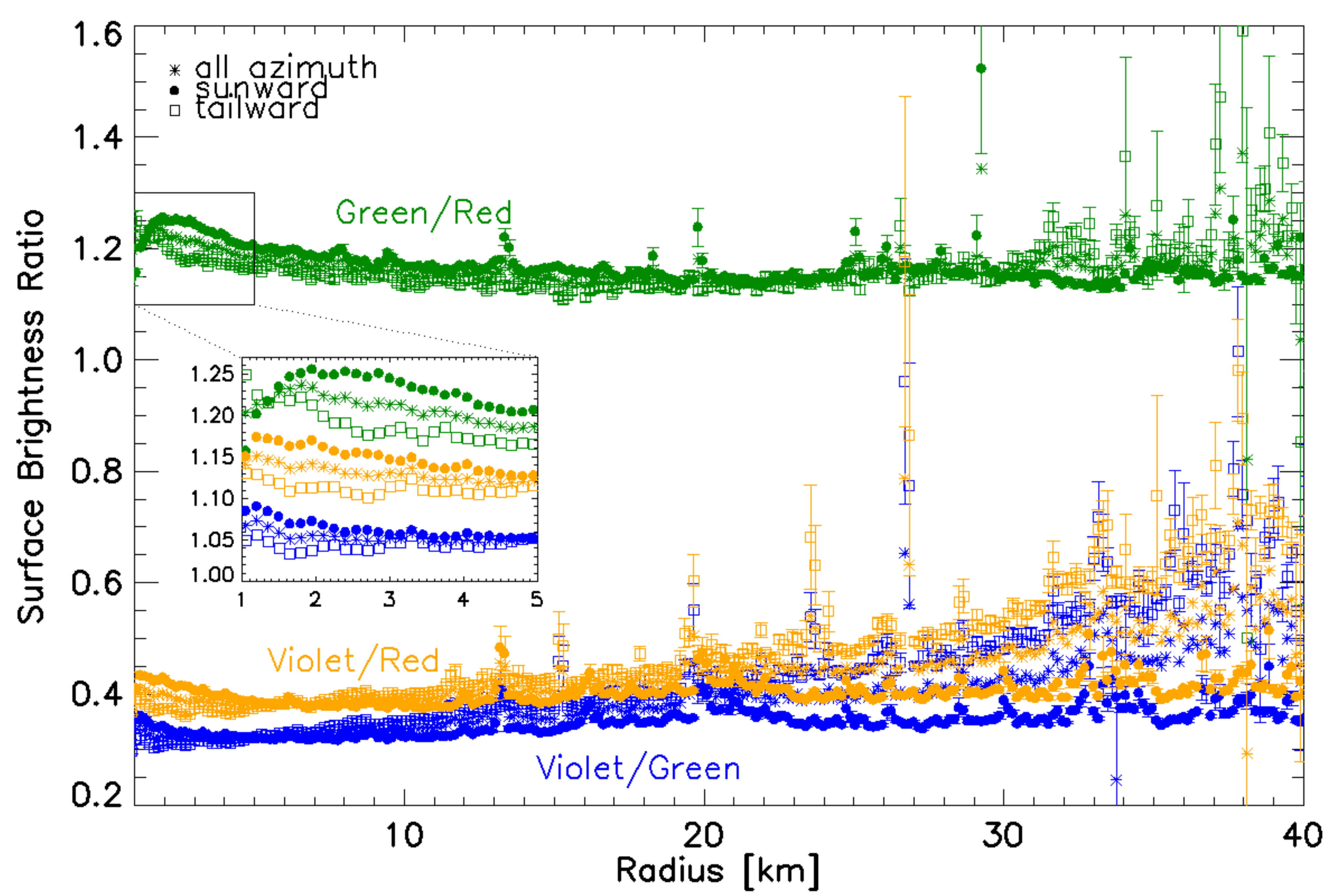}
\caption{Azimuthally averaged radial profiles of color maps 
  Green/Red, Violet/Red and Violet/Green for image 5006065
  (E+8.5 min, FOV 83 km).
  Sunward (0$^\circ<\Theta<180^\circ$) and tailward 
  (180$^\circ<\Theta<360^\circ$) profiles are also shown.
   The inset shows a zoom of the region 1-5 km to emphasize the 
sunward enhancement in all colors. Violet/Red and Violet/Green profiles 
have been shifted by 0.74 and 0.73 respectively to make them visible.
  Error bars represent the statistical error, i. e. the standard deviation 
  divided by the square root of the total number of pixels used to 
  compute the average.
  Very similar profiles are found for image ID 5002027 (E-11 min, FOV 83 km).}
  \label{fig:dust_col_prof}
  \end{figure}

Azimuthally averaged radial profiles of Violet$/$Green, Green$/$Red and Violet$/$Red 
color maps are shown for image ID 5006065 in Fig.\ref{fig:dust_col_prof}
with their sunward ($0^\circ\leq\Theta\leq180^\circ$) and anti-sunward 
($180^\circ\leq\Theta\leq360^\circ$) morphology to point out 
a possible differentiation in the nature 
of the dust. 
 A zoomed inset shows the region 1-5 km to emphasize the 
sunward enhancement in all colors. Violet/Red and Violet/Green 
have been shifted by 0.74 and 0.73 respectively to make them 
visible in the zoom. Error bars represent the statistical error, i. e. the standard deviation 
  divided by the square root of the total number of pixels used for 
  the average. 
  
A significant difference is visible between sunward and tailward profiles 
in all three colors.
The sunward profiles show a consistent enhancement in all three colors
with respect to tailward profiles 
within the first 5 km from the nucleus, where the signal to noise
is high and the statistical errors are small. 
This suggests that within 5 km from the nucleus 
the continuum is bluer in the sunward than in the tailward 
direction. A possible explanation is an enrichment in water ice.
This is strongly supported by the presence of a water ice jet extending to
 5 km from the nucleus, observed in the spectral map acquired 
 with the High Resolution Imager (HRI-IR) 7 minutes after encounter 
 \citep{AHearn2011, Protopapa2014}.
This color variation, if associated with the water ice jet, 
suggests that visible colors are sensitive to the presence of water 
ice in the solid component of the coma. 

At larger distances, beyond 20 km from the nucleus, 
the sunward profiles remain fairly flat, while the tailward profiles 
show a slight increase.
At large radii the statistical errors becomes important because 
the S/N reduces significantly, in particular for Violet/Red and Violet/Green
profiles, where the enhancement is more evident. 
However, these profiles may suggest a variegation of the solid coma at 
these distances, getting bluer in the tailward hemisphere.
 
Only the refractories survive long enough to reach those distances
or possibly big icy particles propelled by a rocket effect 
from sublimation \citep{Kelley2013}. The presence of a larger density of icy particles
in the tailward direction may explain the color profiles differentiation.
However such variations may be also related to other physical processed 
happening in the dust coma such as fragmentation. 

\begin{figure}[h!]
\epsscale{1.2}
\plotone{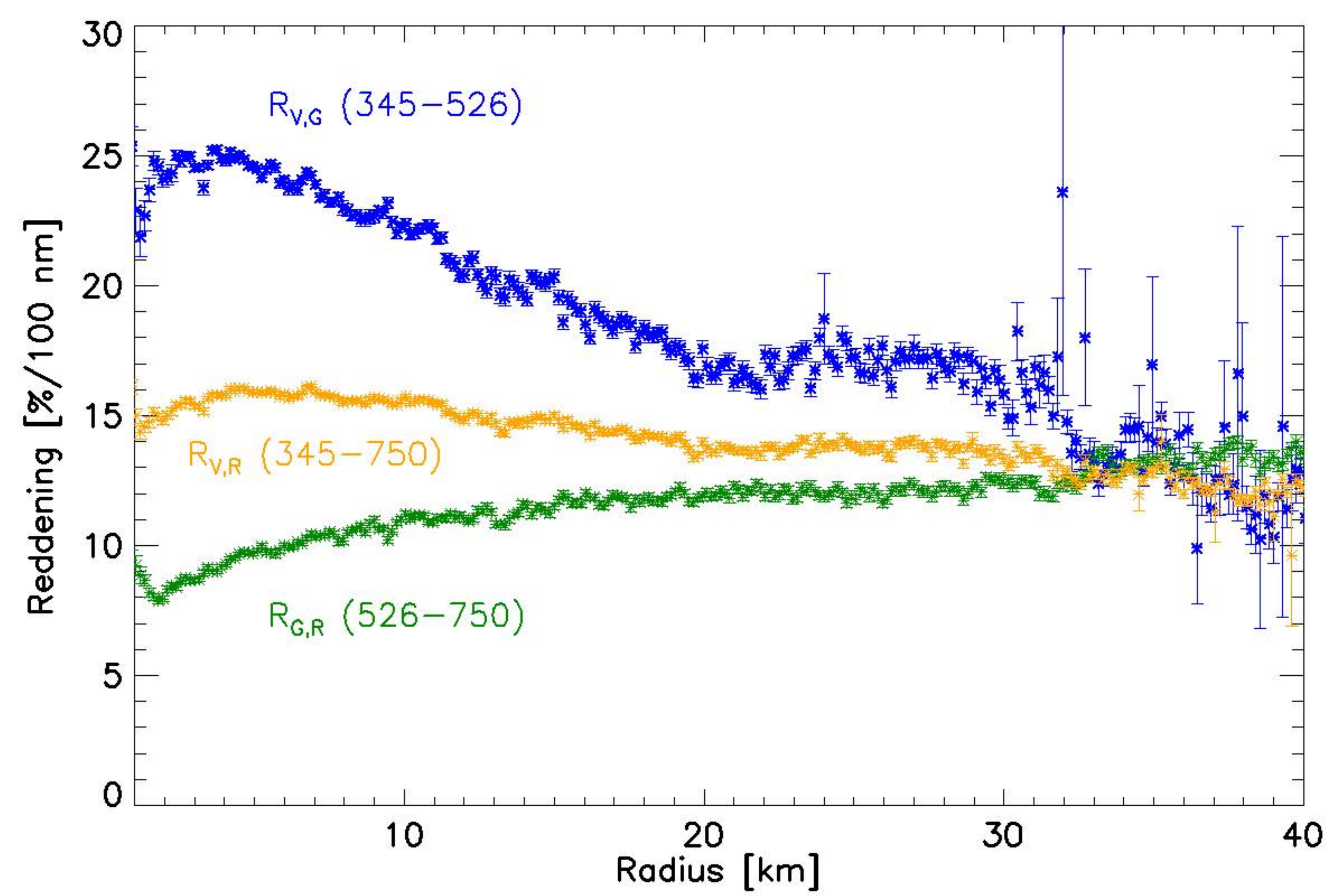}
\caption{Azimuthally averaged radial profiles of the reddening maps
  $R_{V,G}$, $R_{G,R}$ and $R_{V,R}$ for image 5006065.
  Very similar profiles are found for image ID 5002027.}
\label{fig:dust_redd_prof}
\end{figure}

In order to compare the dust colors with other observations of Hartley 2
 we computed the dust reddening maps
using the formula by \citet{Turner1999}:
 \begin{equation}
 R_{1,2}=\frac{{(\frac{I}{F})}_{2}-{(\frac{I}{F})}_{1}}{\lambda_{2}-\lambda_{1}}\frac{20000}{{(\frac{I}{F})}_{1}+{(\frac{I}{F})}_{2}}
\end{equation}
where $1$ and $2$ stands for the shorter and longer 
wavelength respectively among $V$ (Violet), $G$ (Green) and $R$ (Red).
Fig. \ref{fig:dust_redd_prof} shows the azimuthally averaged radial profiles 
of the three reddening maps for image ID 5006065.
$R_{V,G}$ is about 25\%/100 nm close of the nucleus, 
but decreases to about 12\%/100 nm at 40 km.
Our results are consistent with the observations by \citet{Lara2011}, 
who showed various reddening profiles in the range 415--693 nm that indicate 
a decrease of the reddening with distance, as well as variations
with direction in the coma. 
\citet{Knight2013} found reddening lower than $10\%/$100 nm in the wavelength
range 345--526~nm within an aperture radius of thousand km.

$R_{G,R}$ is about 9\%/100 nm in the 
vicinity of the nucleus but it increases slightly reaching 13\%/100 
nm at 40 km. This profile stays much flatter than $R_{V,G}$.
The profiles suggest that the spectrum of the dust in the vicinity 
of the nucleus is similar to the spectrum 
of the comet's nucleus \citep{Li2013} and consistent with the spectral 
properties of carbonaceous material, which is steeper 
at bluer wavelengths and flatter at visible wavelengths \citep{Jewitt1986}.
At about 40 km the spectrum of dust coma flattens, and the 
 slopes in the blue and visible regions are consistent with 
the results found by \citet{Knight2013} for the dust coma at larger distances
from the nucleus.
This could be interpreted as a change of scattering properties of the
solid coma
due for example to fragmentation or sublimation of one component (organics).
Note that $R_{1,2}$
is a function of wavelength, and that differences in $R$ computed using
two different sets of filters (e.g., V-G, G-R) should not necessarily
be taken as a change in the spectral slope.  We can make general
comparisons between the filter sets, and to the literature, but small
differences should not be considered significant.

\subsection{Continuum removal}
\label{sec.OHcont}
\begin{figure}[h!]
\epsscale{1.2}
\plotone{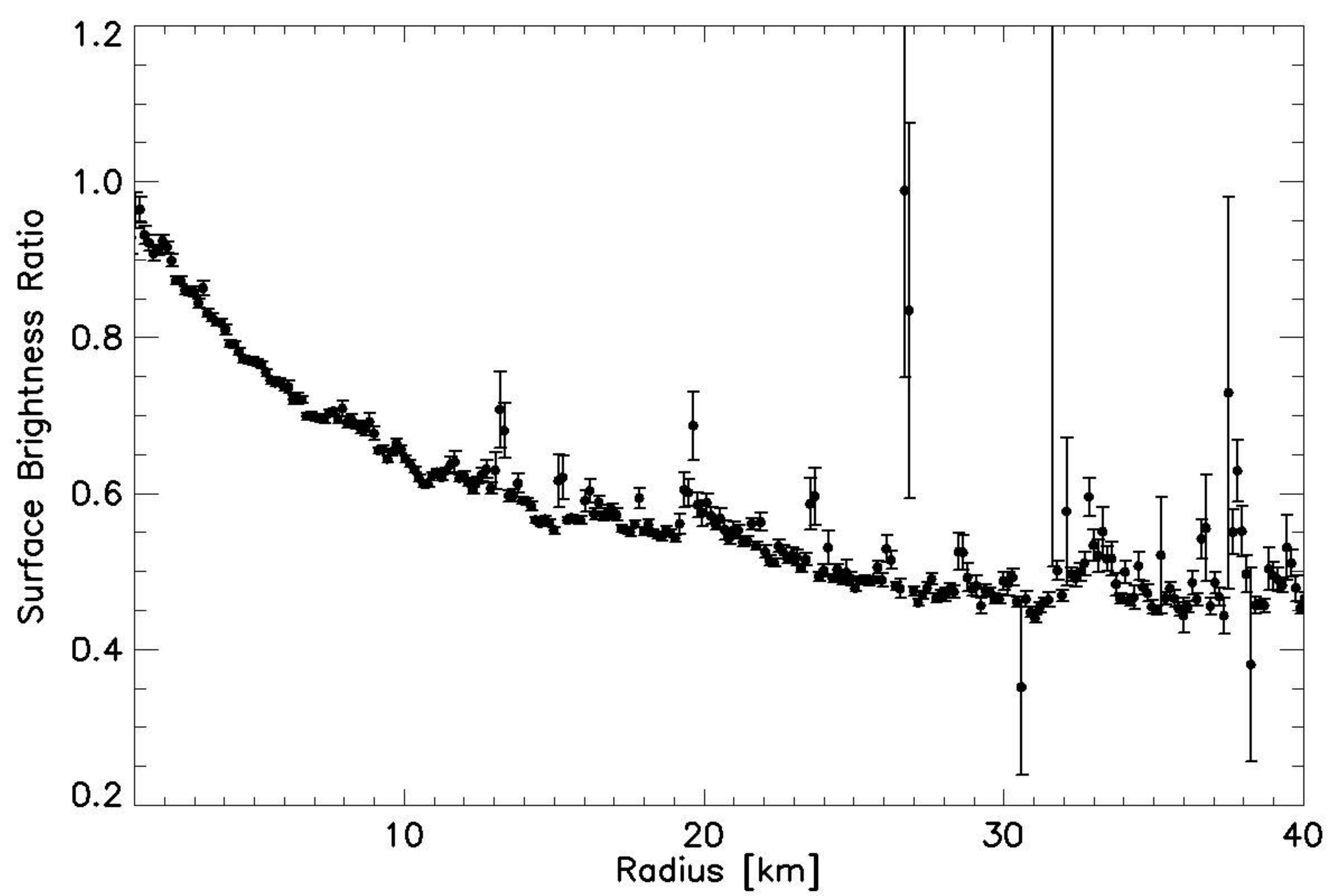}
  \caption{\small Ratio between the resulting subtracted continuum
  and the OH original signal azimuthally averaged
  radial profile for OH image 5006065.
  Very similar profile is found for image ID 5002027.}
  \label{fig:conttooh}
\end{figure}

To remove the continuum from the OH images IDs 5002027 and 5006065,
we assumed that for each pixel in the frame the solid coma 
has a reflectance spectrum approximated 
by two straight lines respectively in the wavelenghts ranges
345-526 nm (Violet-Green) and 526-750 nm (Green-Red). 
We then extrapolated the slope of Violet-Green line
to the effective wavelength
of the OH filter (309.48 nm) for every pixel.
The resulting subtracted continuum is on average 45$\%$ of the total 
surface brightness, but in the very vicinity of the nucleus it reaches 
$\sim95\%$ of the original signal (Fig. \ref{fig:conttooh}).

For the remaining 
151 OH images, the contemporaneous continuum observations were 
mostly acquired with the Clear1 filter only and no colors or reddening studies can be derived for those images.
We therefore used the color information acquired around CA and assumed that the solid coma spectrum remained 
 constant in time during the whole encounter.  
 
For this extrapolation we computed the ratio of the continuum in the 
OH filter and the signal in each color and Clear1 filter for the images 
acquired at CA.
The azimuthally averaged profiles of all resulting OH/color filter ratios 
are fairly flat and uniform.  
Assuming that the dust colors did not change
significantly over the considered period, 
we used the resistant mean of the ratio 
between the OH profile and each continuum filter profile as the
{\it continuum removal factor} for that filter.
The computed continuum removal factors are: 0.237 for Clear1 filter, 0.615 for Violet filter, 
0.201 for Green filter and 0.235 for the Red filter. For comparison, the grey continuum
removal factors \citep{Klaasen2013} computed assuming that the comet has a solar-like spectrum
(0.443 for Clear1, 0.668 for Violet, 
0.337 for Green and 0.497 for Red) are
significantly different, suggesting that the assumption of a solar spectrum, 
with no reddening, would have significantly overestimated 
the continuum contribution to the flux measured with the OH filter.

Using the Clear1 observations for the continuum  removal
is problematic since it is a broad filter \citep[see][]{Hampton2005}, 
and the flux within
its bandpass inevitably contains gas emissions such as 
CN, C$_2$, C$_3$, NH, NH$_2$. 
To evaluate this contamination of the Clear1 signal
we generated a ``synthetic'' Clear1 image 
using the two-straight line spectrum derived from the color images at CA
and the Clear1 filter bandpass.
The comparison between the observed Clear1 brightness and the synthetic image  
provided an estimate of the gas contamination of the Clear1 filter.
It ranges between 9\% in the vicinity of the nucleus to about 12\% at the 
edges of the field of view, slightly increasing with 
nucleocentric distance. 
However, using the {\it continuum removal factors} computed
as described above we indirectly subtract this contribution from Clear1
filter allowing a more precise continuum subtraction with the sole
assumption that the colors of the dust remained the same as at CA.

\subsection{Uncertainties}

MRI-Vis calibration is expected to produce errors of less than 10\%
\citep{Klaasen2013} on the total surface brightness.
The continuum removal represents the largest uncertainty in measuring 
the ``pure'' gas surface brightness.

The original S/N of OH images was typically $\sim$ 5 reaching $\sim$ 13 in the 
vicinity of the nucleus. Similar values applied to Violet images. While 
Green observations had generally higher S/N of about 10 with maximum of 22
in the vicinity of the nucleus. 
Assuming that each image had a calibration error of 10\% and propagating 
the errors for the linear extrapolation, 
the resulting pure OH gas surface brightness images have
an estimated error of about 50\%.

\section{OH FLUORESCENCE EMISSION}

  \label{sec:OHfluorescence}

    \setcitestyle{citesep={;}}
Once the continuum has been  removed, the \textquotedblleft pure gas\textquotedblright \hspace{0.5mm}image can be
retrieved and converted into OH
column densities $N_{OH}$, 
i.e. the number of molecules along the line of sight in cm$^{-2}$, through 
the formula:

\begin{equation}
 N_{OH}=\displaystyle\frac{4\pi \Delta^2 \Omega d\lambda}{p^2g'_{OH}}F_{OH}
 \label{eq:colden}
\end{equation}
where $\Delta$ is the distance spacecraft-comet; $\Omega$ is the solid angle of a pixel; 
$d\lambda$ is the bandwidth; $p$ is the pixel scale of MRI; $F_{OH}$ is the 
radiance of the pure gas measured in W m$^{-2}\mu$m$^{-1}$ sr$^{-1}$
and $g'_{OH}$ is the fluorescence efficiency, or $g$-factor, of the OH band
as measured through the MRI-OH filter at the heliocentric distance and velocity 
of Hartley 2.

The $g$-factor describes the efficiency of the molecules in emitting light, 
and is defined \citep{Chamberlain1987} in cgs units at 1 au for a generic molecule and a single
emission line $\lambda$ by:

\begin{equation}
 g(\lambda) =\frac{\pi e^2}{mc^2}\lambda^2 f_\lambda\pi F_{\odot} \tilde{\omega}
 \label{eq:gfactor}
\end{equation}
where $e$ and $m$ are the charge and mass of the electron and $c$ is the 
speed of light,
$f_\lambda$ is the absorption oscillator strength
of the molecule, $\pi F_\odot$ is the solar flux per unit wavelength at 1 au, 
$\tilde{\omega}$ is the relative Einstein coefficient 
for the given line.
This has to be scaled for the actual heliocentric distance of the comet at the observation time.

\citet{Swings1941} pointed out that the fluorescence efficiency of some 
molecules also depends on the heliocentric velocity of the comet.
This is because the visible region of the solar spectrum contains strong Fraunhofer absorption lines.  The relative motion between the comet and  the Sun causes a Doppler shift of the solar spectrum at the comet, affecting the excitation of OH as the lines move in and out of resonance. A change in heliocentric velocity leads to observable differences in the structure of 
the bands and is particularly important for OH, CN and NH. 
\citet{Schleicher_AHearn1988} computed the fluorescence efficiency 
for the NUV band of OH for a wide range
of heliocentric distances and velocities, and showed that it varies up 
to a factor of 5.
\begin{figure}[ht!]
\epsscale{1.2}
\plotone{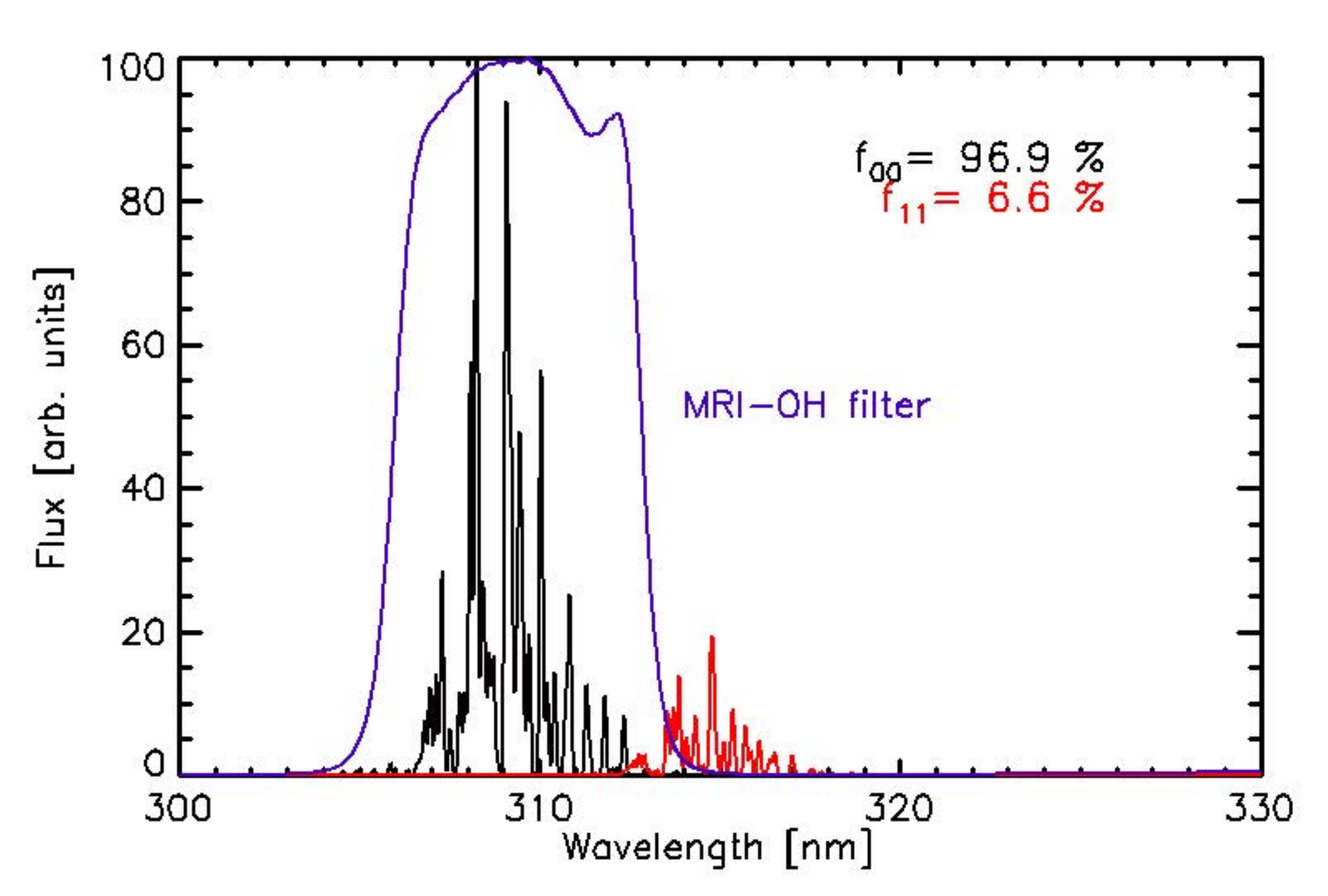}
    \caption{\small Spectrum of OH fluorescence (0,0) (black) and (1,1) 
    (red) bands obtained using LIFBASE software and 
    the fluorescence levels population distribution 
    (D.~Schleicher, priv. comm.) at the heliocentric distance
    and velocity of Hartley 2.
    Violet curve shows the MRI-OH bandpass filter weighted
    for the CCD QE and mirror reflectivity.}
    \label{fig:band_fraction}
\end{figure}

\begin{figure}[htb!]
\epsscale{1.2}
\plotone{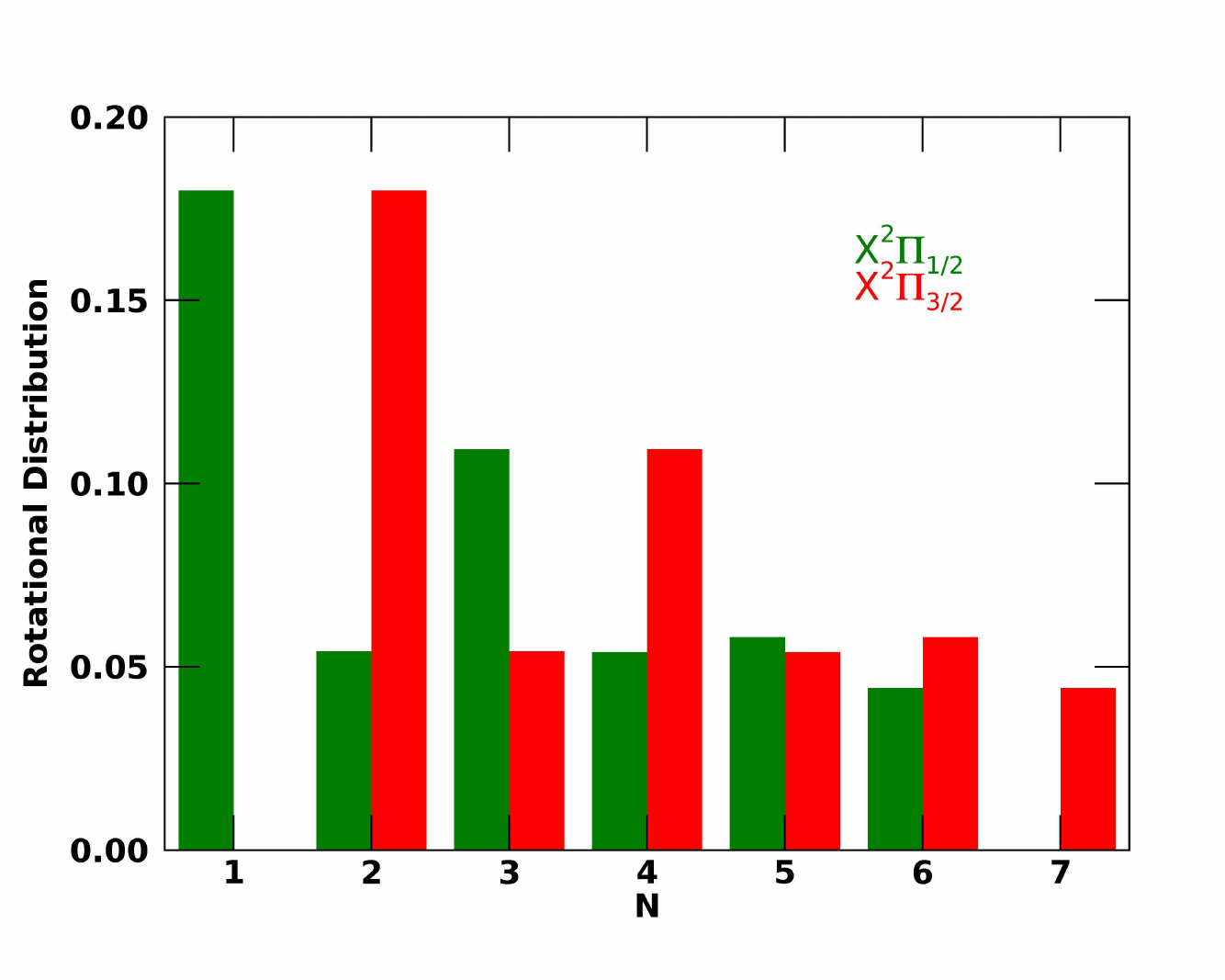}
    \caption{Rotational population distribution of the ground 
    vibrational state for pure fluorescence calculations at the time
    of EPOXI encounter with Hartley 2. (Schleicher, priv. comm.)}
    \label{fig:Schleicher}
\end{figure} 

\begin{figure*}[!t]
\gridline{
\fig{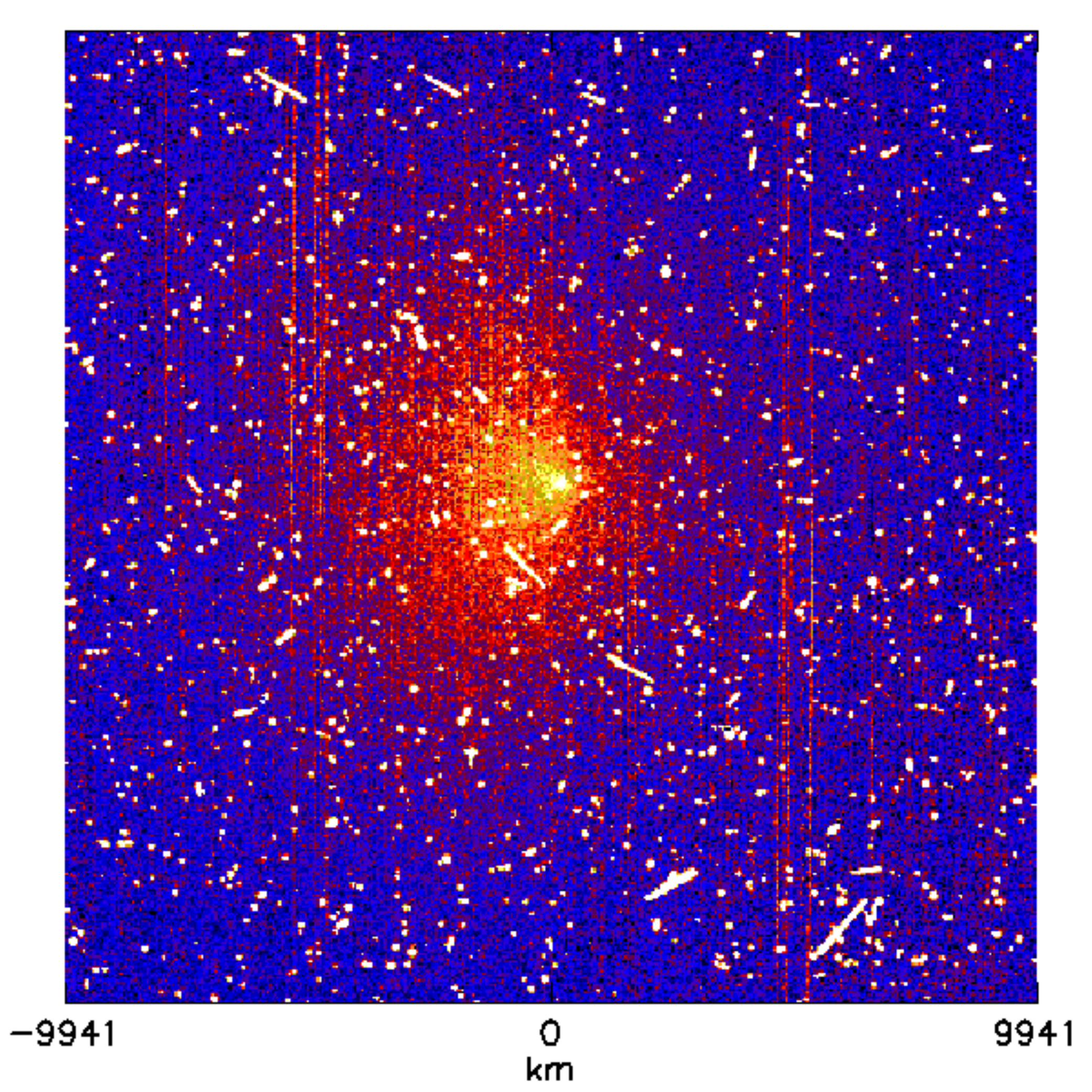}
{0.33\textwidth}{(a)}
\fig{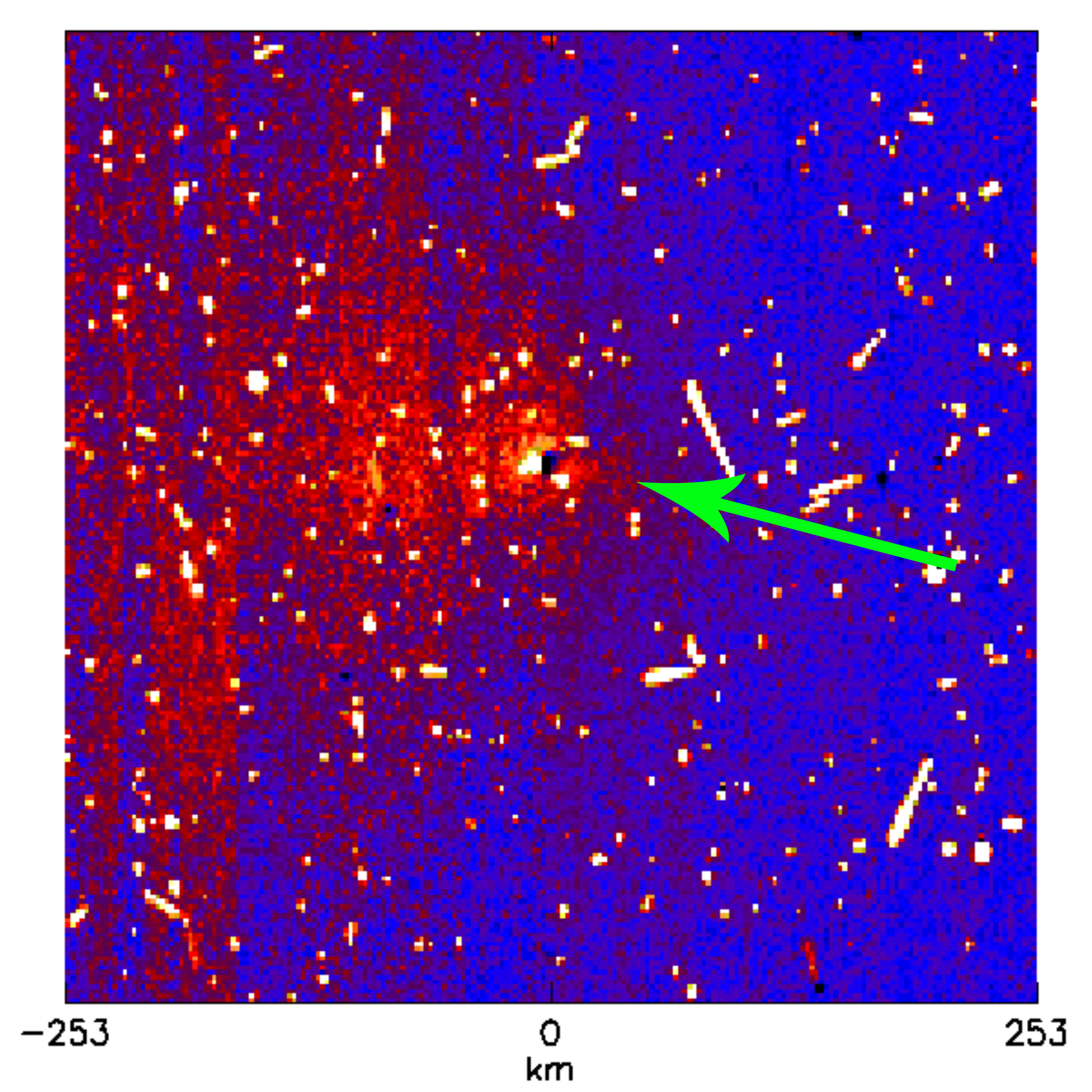}
{0.33\textwidth}{(b)}
\fig{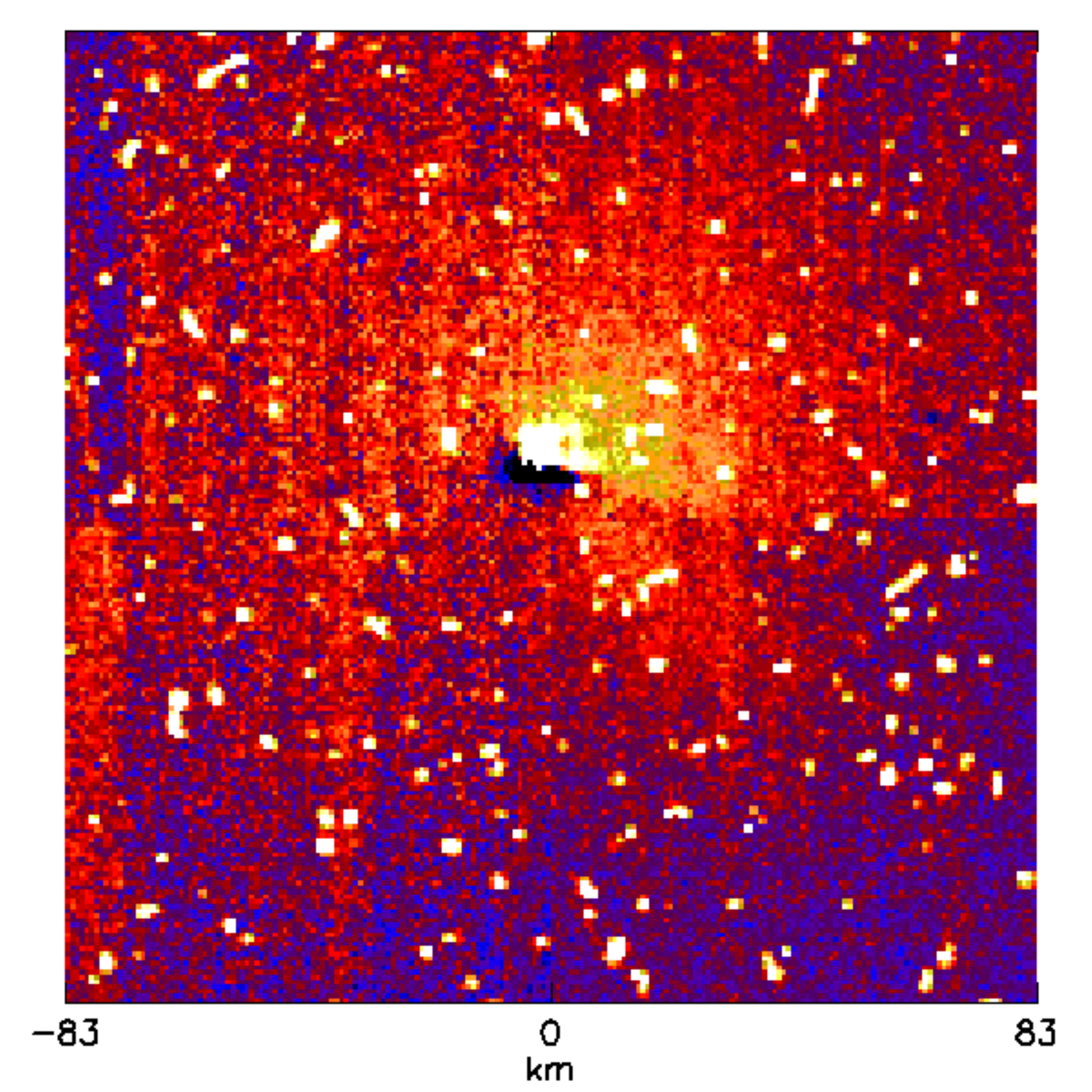}
{0.33\textwidth}{(c)}}
  \caption{OH column density maps pre-CA. Sun is on the right side.\\
  (a) DOY 304 at 22:38 UTC; FOV 20000 km; spatial scale
39 km pix$^{-1}$; linear color scale with range $[10^{12},3 \times 10^{13}]$ mol cm$^{-2}$ \\
(b) DOY 308 at 09:22 UTC; FOV 500 km; spatial scale 2 km pix$^{-1}$; 
linear color scale with range $[10^{12},7 \times 10^{13}]$ mol cm$^{-2}$\\
(c) DOY 308 at 12:16 UTC; FOV 188 km; spatial scale 0.7 km pix$^{-1}$; 
linear color scale with range $[10^{12},7 \times 10^{13}]$ mol cm$^{-2}$}
\label{fig:CDmaps_pre}
\end{figure*}

\begin{figure*}[!t]
 \gridline{
 \fig{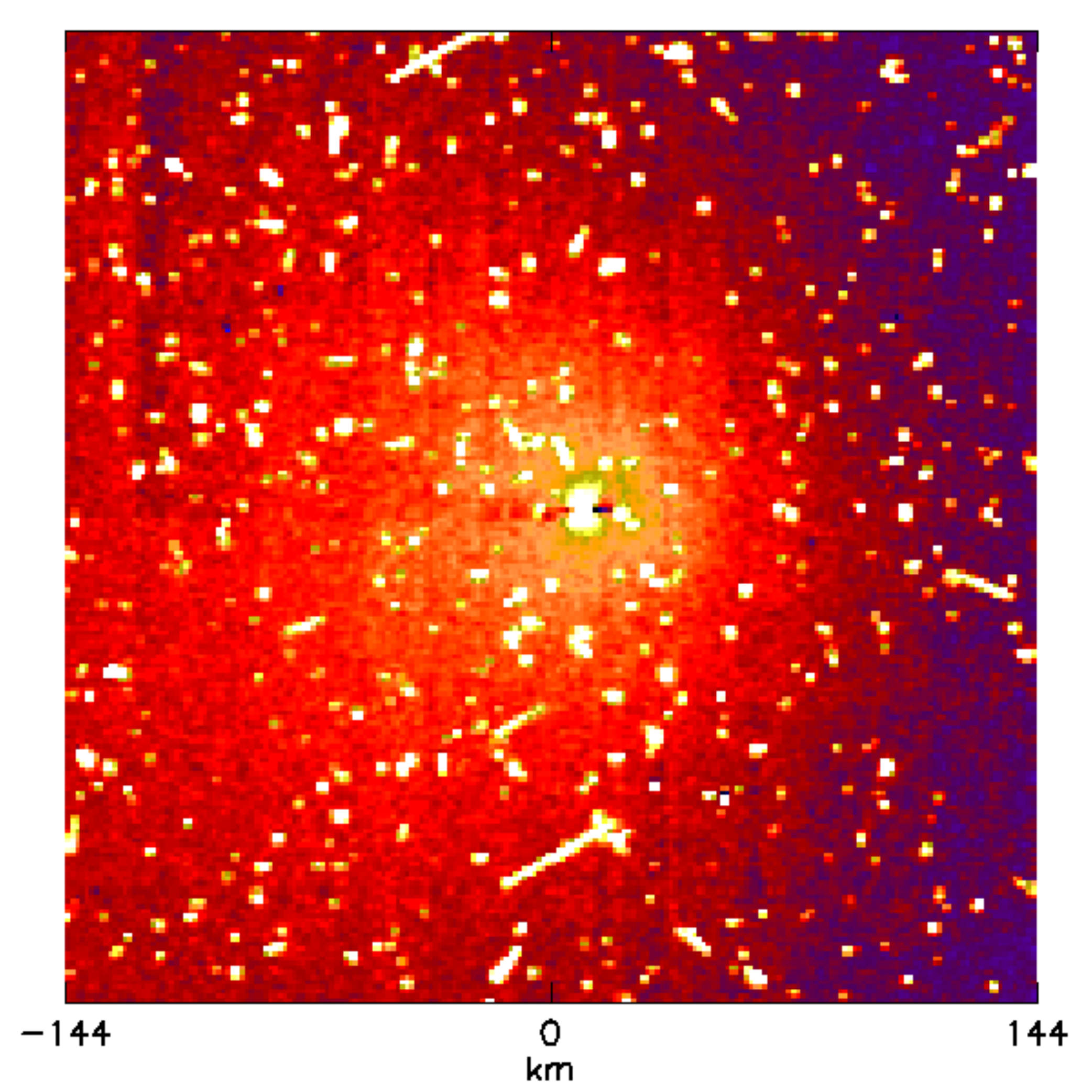}
 {0.33\textwidth}{(a)}
 \fig{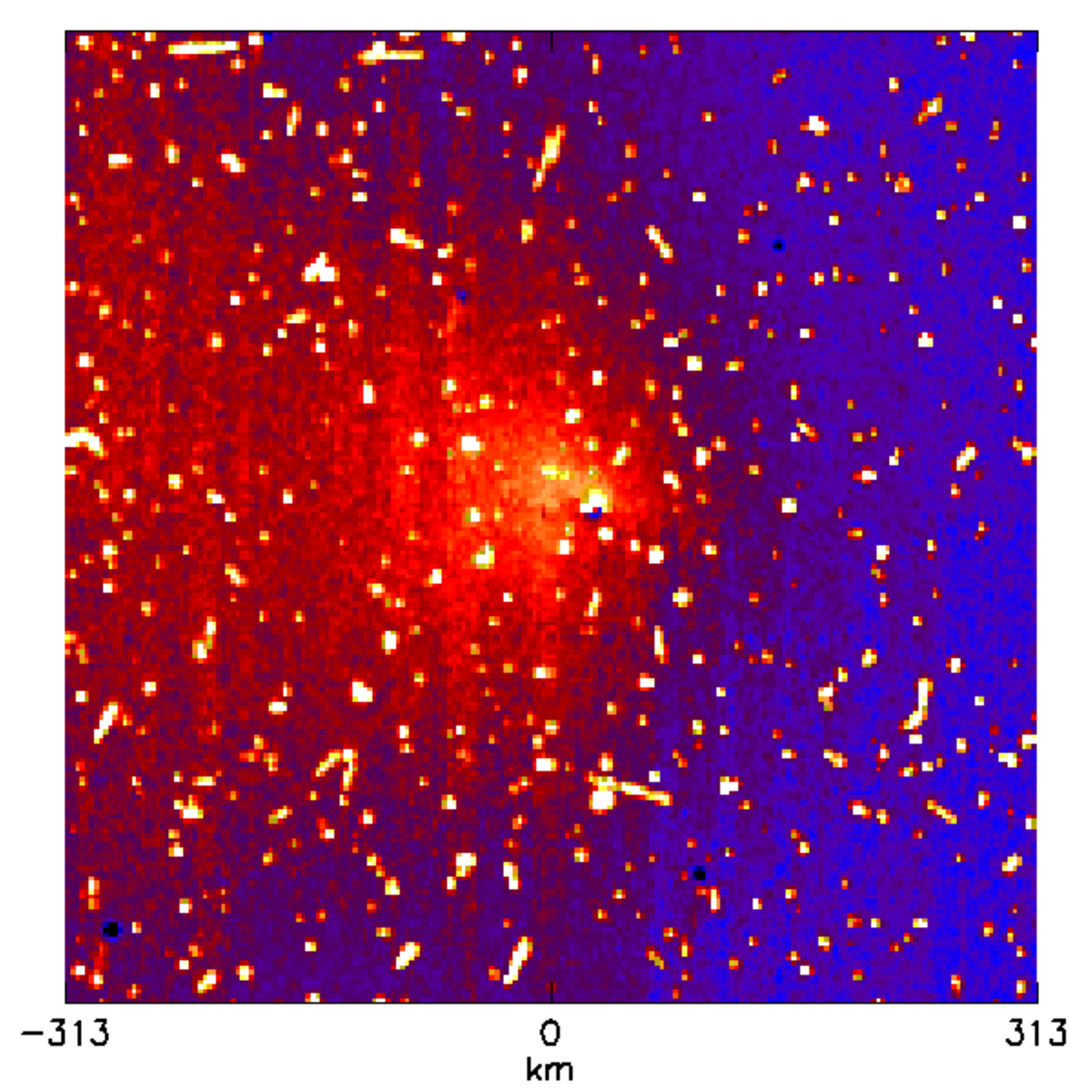}
 {0.33\textwidth}{(b)}
 \fig{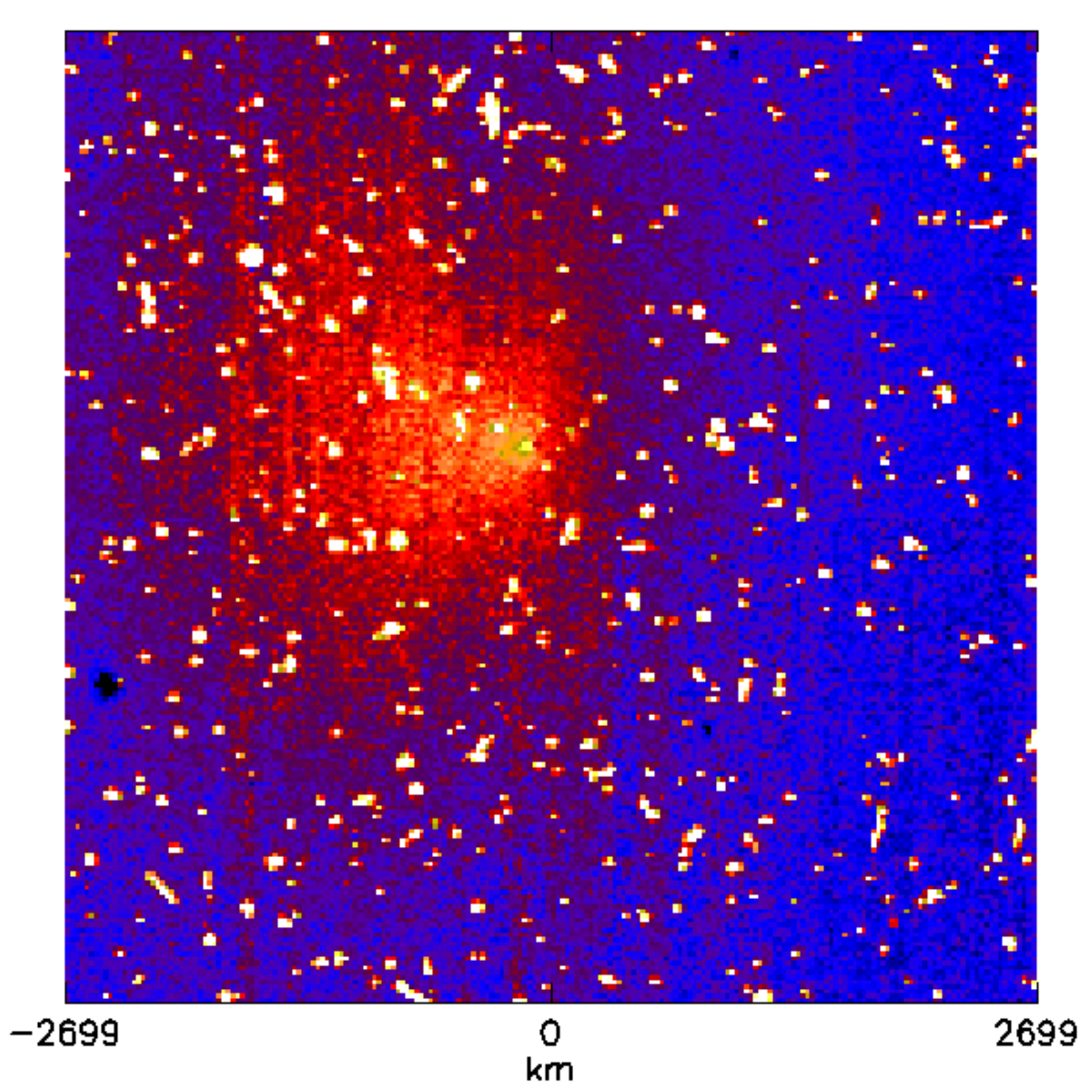}
 {0.33\textwidth}{(c)}}
\caption{OH column density maps post-CA. Sun is on the right side. \\
(a) DOY 308 at 16:31 UTC; FOV 300 km; spatial scale 1.2 km pix$^{-1}$; 
linear color scale with range $[10^{12},7 \times 10^{13}]$ mol cm$^{-2}$ \\
(b) DOY 308 at 19:31 UTC; FOV 640 km; spatial scale 2.5 km pix$^{-1}$; 
linear color scale with range $[10^{12},7 \times 10^{13}]$ mol cm$^{-2}$\\
(c) DOY 310 at 13:46 UTC; FOV 5500 km; spatial scale 21 km pix$^{-1}$; 
linear color scale with range $[10^{12},4 \times 10^{13}]$ mol cm$^{-2}$}
\label{fig:CDmaps_post}
\end{figure*}

\begin{figure*}[!t]
\gridline{
\fig{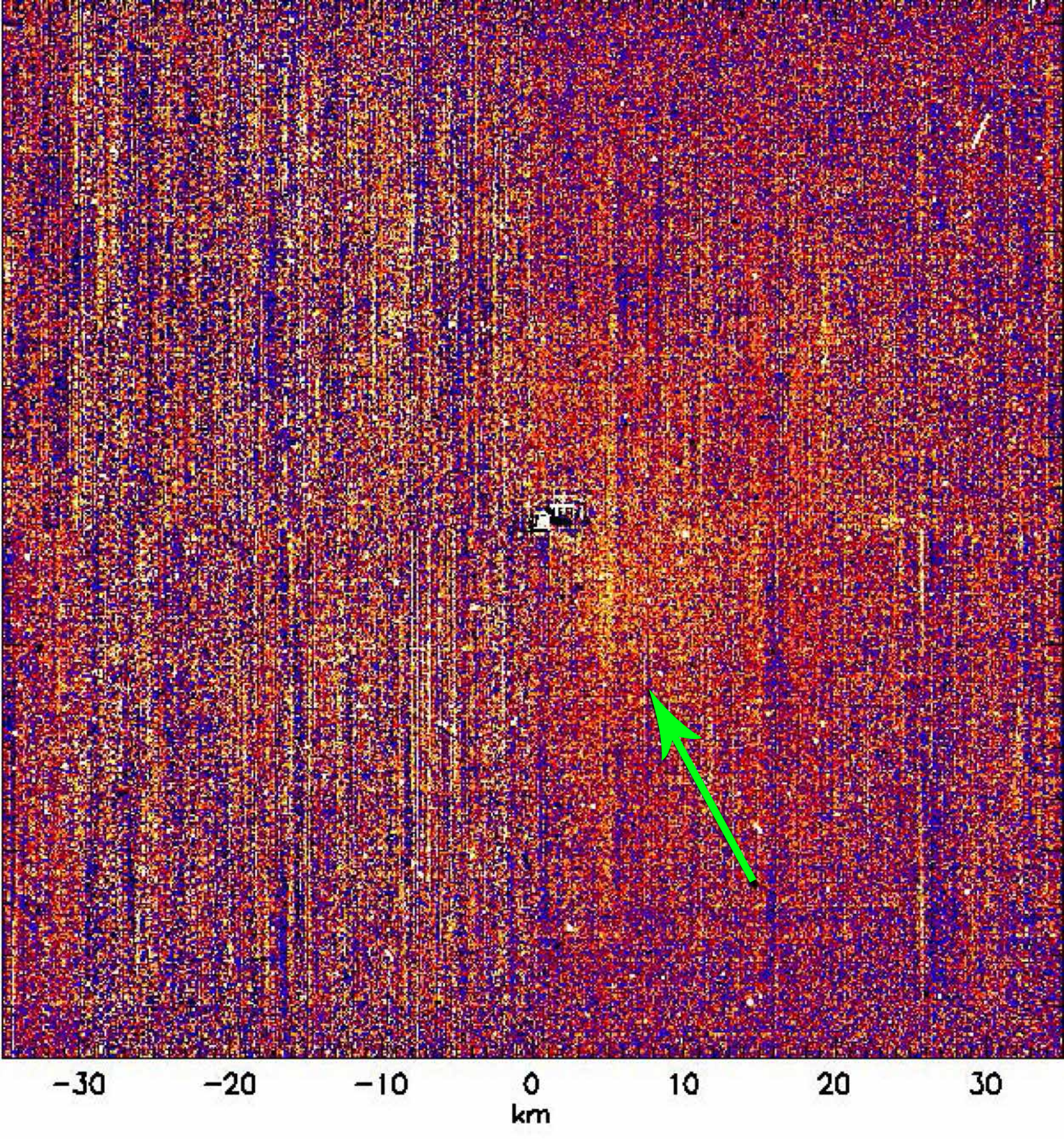}{0.5\textwidth}{(a)}
\fig{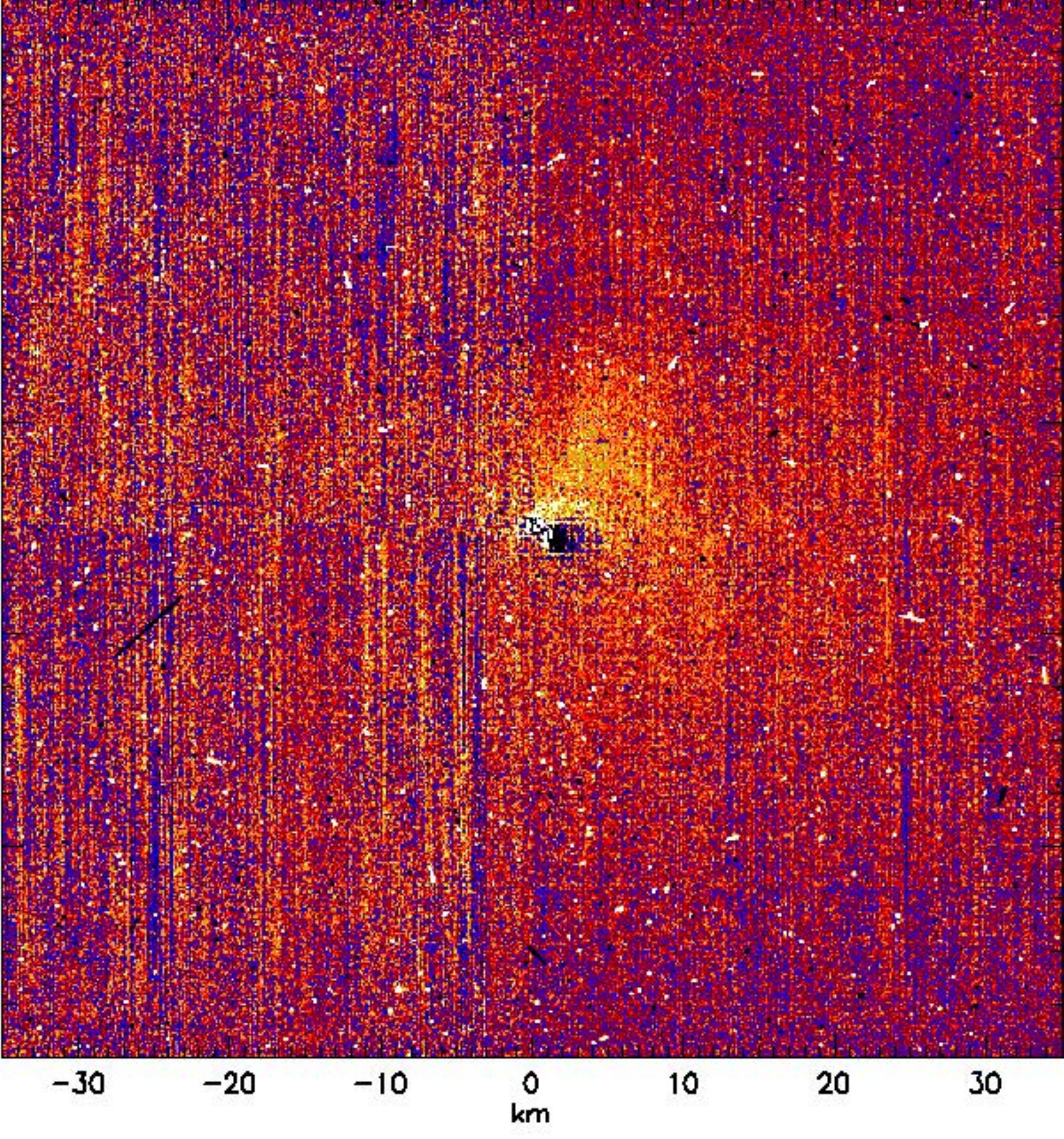}{0.5\textwidth}{(b)}}
\gridline{
\fig{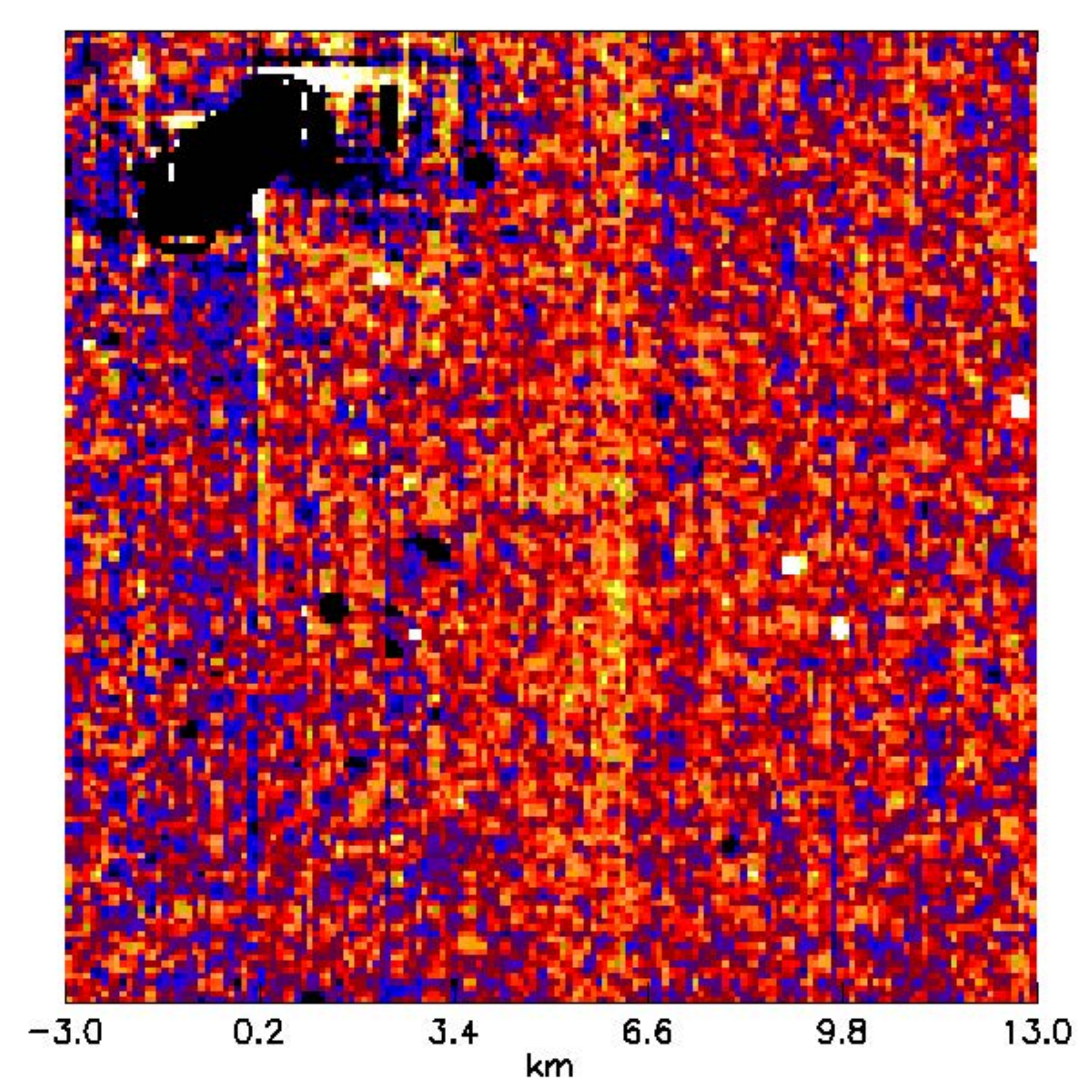}{0.45\textwidth}{(c)}
\fig{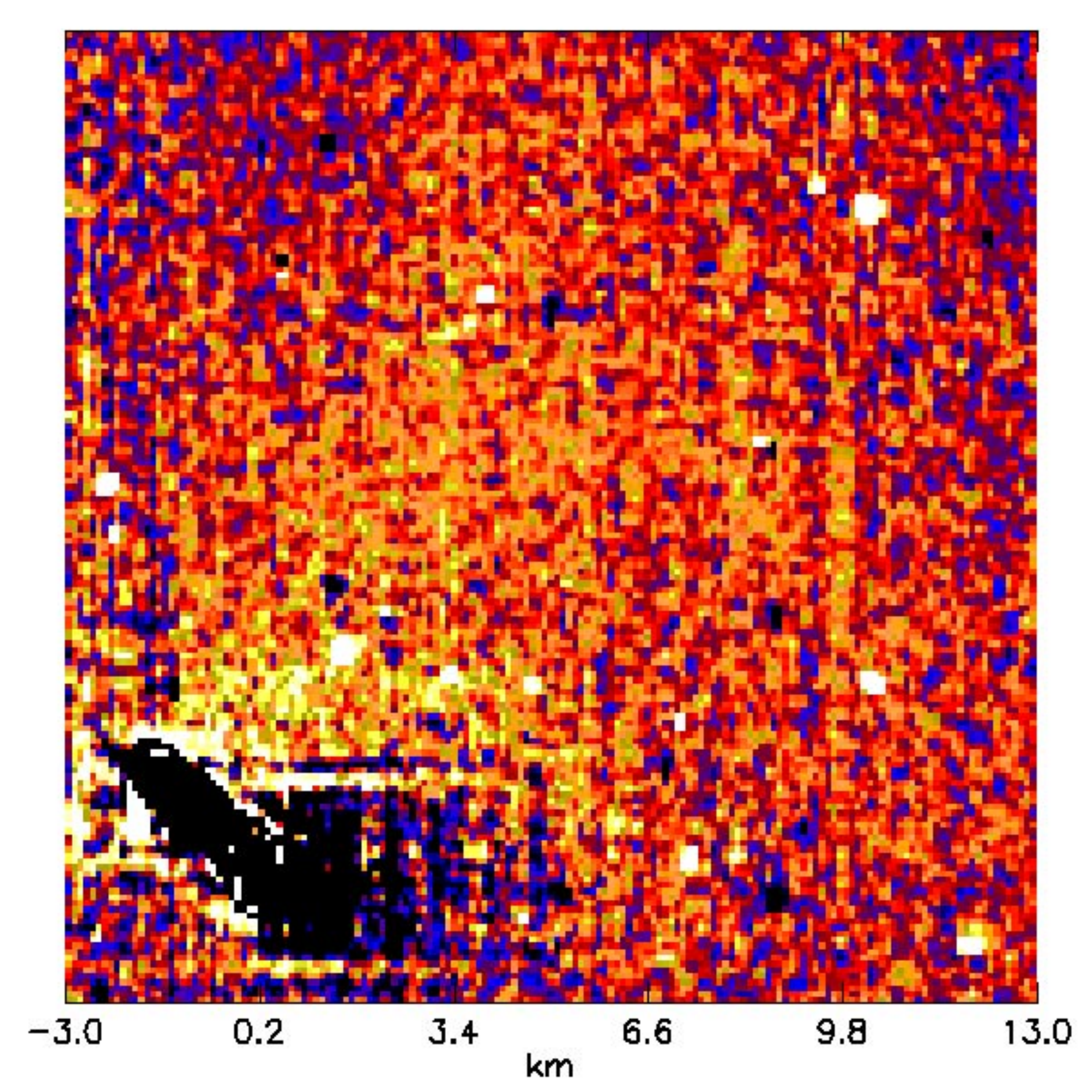}{0.45\textwidth}{(d)}}
\caption{OH column density maps at CA.   
(a) image ID 5002027; DOY 308 at 13:48 UT; spacecraft distance 8303 km; spatial scale 83 m pix$^{-1}$;
linear color scale with range $[10^{12},10^{14}]$ mol cm$^{-2}$
(b) image ID 5006065; DOY 308 at 14:09 UT; spacecraft distance 8256 km; spatial scale 82.6 m pix$^{-1}$. 
linear color scale with range $[10^{12},10^{14}]$ mol cm$^{-2}$.
 (c) and (d) panels show an enlargement of the (a) and (b) panels respectively 
in the near-nucleus region to better highlight the nucleus projection and the 
spatial distribution of OH.}
\label{fig:OHmaps}
\end{figure*}

\begin{figure}[htb!]
\epsscale{2.5}
\plottwo{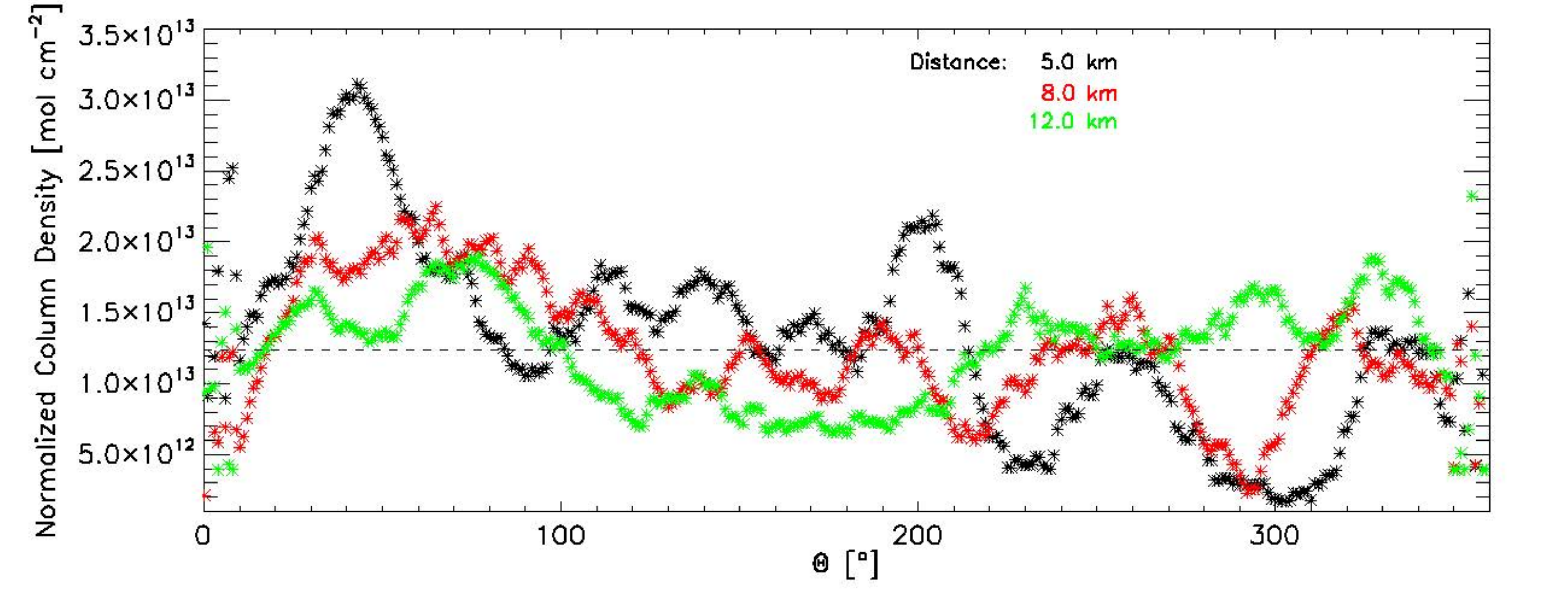}
    {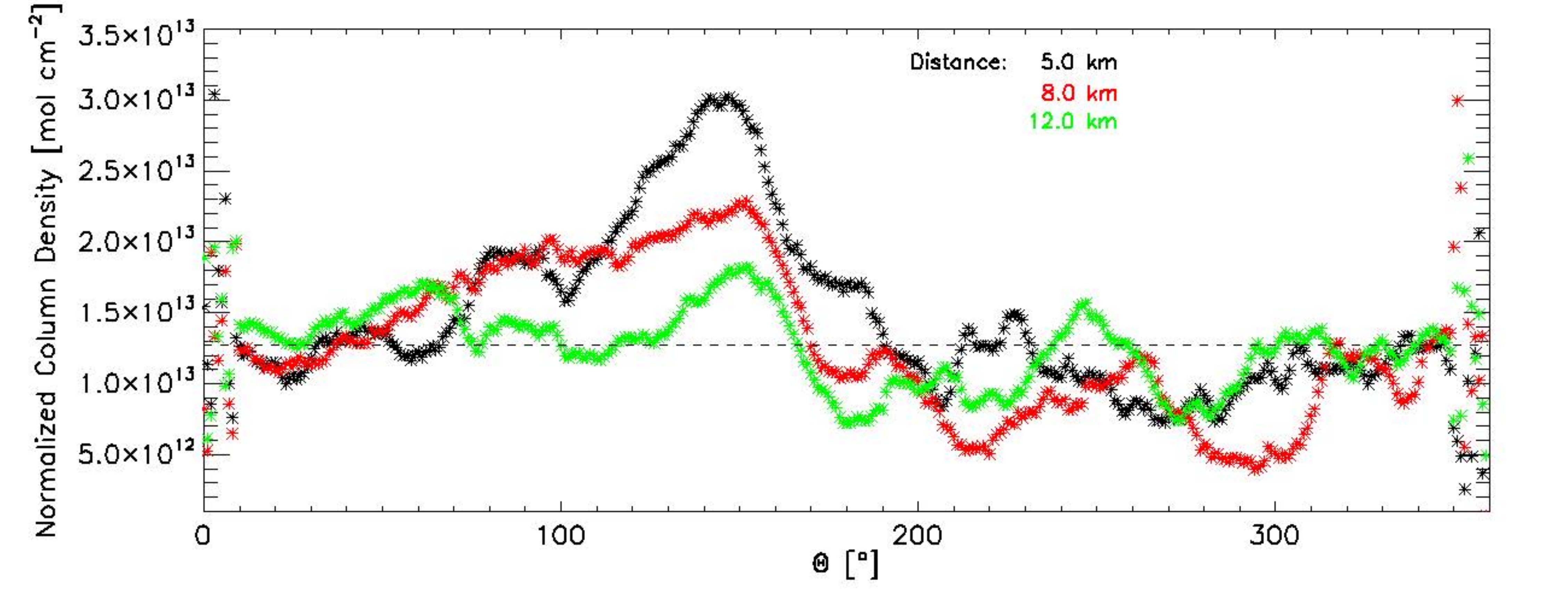}
    \caption{Azimuthal profiles of the OH column density maps in Fig.\ref{fig:OHmaps}: 
    pre-CA image ID 5002027 (top panel) and post-CA image ID 5006065 (bottom panel).
    Profiles are computed as averages over 1-km width rings centered on the nucleus 
    with radii of 5 (black), 8 (red), and 12 (green) km.}
    \label{fig:OHazprof}
\end{figure} 

The OH molecules, excited by solar photons to the first electronic state $A^2\it{\Sigma}^+$,
mainly populate the lowest rotational states in the first two vibrational 
levels $\nu=0$ and $\nu=1$ \citep{Schleicher_AHearn1988}. Following 
the selection rules, OH radicals in these states decay mainly into the vibrational levels 
$\nu=0$ and $\nu=1$ of the ground electronic state, producing 3 bands: 
the (0,0) band, centered at 308.5~nm, the (1,1) band, centered at 314.3~nm, 
and the (1,0) band centered at 283~nm.
The (1,0) band is completely outside the MRI-OH bandpass, 
and to evaluate how much of the (0,0) and (1,1) bands falls within the
MRI filter bandpass, we generated a synthetic OH spectrum. 
For this we use a level population distribution calculated 
for fluorescence excitation only
(D. Schleicher, priv. comm.) at the time of EPOXI encounter
with
Hartley~2, i.e. for a heliocentric distance ($r_h=1.064$ au) and 
a heliocentric
velocity (2.13 km s$^{-1}$) and for gas moving 1 km s$^{-1}$ towards the Sun.
This calculation predicts that the first three vibrational states 
are populated, with fractions of 75.4\%, 23.2\% and 1.4\% respectively.
The ground vibrational state population is in turn distributed in the 
first 7 rotational levels (Fig. \ref{fig:Schleicher}).

We then used the LIFBASE software \citep{Luque1999} to generate the resulting 
emission spectra $S_{00}$ and $S_{11}$.

We weighted this synthetic spectra $S_{00}$ and $S_{11}$ 
with the MRI CCD quantum efficiency
$Q(\lambda)$, mirror reflectivity $R(\lambda)$, and 
transmission of the OH filter $T(\lambda)$ to calculate the 
fraction $f_b$ of the band flux which is measured by the filter:
\begin{equation}
 f_{b}=\frac{\int{S_{b}Q(\lambda)R(\lambda)T(\lambda)}}{\int{Q(\lambda)R(\lambda)T(\lambda)}}
\label{eq:convolution}
 \end{equation}
where $b$ can denote either (0,0) or (1,1).
The MRI-OH filter has a bandpass of 6.2 nm \citep{Hampton2005} which
includes most of the (0,0) band (96.9$\%$) and a small 
fraction of the (1,1) band (6.6$\%$) 
(see Fig. \ref{fig:band_fraction}).

We used these fractions to calculate the effective fluorescence 
efficiency $g'_{OH}$ as seen through the MRI OH filter as:
\begin{equation}
 g'_{OH}=\frac{g_{00}(\dot{r}_h)f_{00}+g_{11}(\dot{r}_h)f_{11}}{r_h^2}
 \label{eq:goh}
\end{equation}
where $g_{00}(\dot{r}_h)$ and $g_{11}(\dot{r}_h)$ are the $g$-factors 
of the (0,0) and (1,1) bands respectively \citep{Schleicher_AHearn1988}, 
at the heliocentric distance ($r_h$) and
velocity ($\dot{r}_h$) of Hartley 2.

\section{Results: OH in the coma of 103P/Hartley 2}

\subsection{The Spatial Distribution of OH}
\setcitestyle{citesep={;}}

We generated OH column density maps for the 153 OH images acquired between the 
day of the perihelion 
up to 10 days afterwards using Eq. \ref{eq:colden} and the effective $g$-factor 
in Eq. \ref{eq:goh}. 
We selected eight examples of column density maps 
from the full dataset to 
show the spatial distribution of OH gas and structures in the coma
as representative for three distinct epochs.
Fig. \ref{fig:CDmaps_pre} shows column densities obtained before CA,
Fig. \ref{fig:CDmaps_post} shows three examples acquired after CA, 
while Fig. \ref{fig:OHmaps} shows the OH column density maps of two most resolved images
acquired during CA (IDs 502027 and 5006065, see Table \ref{tab.images}). 
In all images the Sun is located on the right side. 

We first look at the images pre-CA and post-CA and then we
treat differently the two images in Fig. \ref{fig:OHmaps} since they 
have a much smaller scale.
A clear anti-sunward enhancement of the OH coma is visible in most of the 
approach images (DOY 301 -- 308). This enhancement has a very sharp fan-like 
shape in the anti-sunward direction (Fig. \ref{fig:CDmaps_pre}a).
As the spatial resolution increases, a radial jet becomes visible in the 
sunward direction
about 4 hours before CA (see arrow in Fig.~\ref{fig:CDmaps_pre}b) and 
reaches its maximum brightness at DOY 308 $\sim$ 11:23 UTC.
In the following observations, towards the approach to the comet, 
the OH column density distribution becomes more isotropic with a
slight enhancement in the sunward direction (Fig. \ref{fig:CDmaps_pre}c).
After CA, the distribution remains isotropic 
for very short time (Fig. \ref{fig:CDmaps_post}a), 
after which again an enhancement in the anti-sunward direction becomes visible (Fig. \ref{fig:CDmaps_post}b) 
slightly more toward the left top corner (Fig. \ref{fig:CDmaps_post}c)
with respect to the pre-CA images.

Although Hartley 2 has been studied in great detail from the ground and from space, its 
rotational state is not yet fully understood due to its very complex rotation.
\citep{Belton2013, Knight2015}.
Therefore, it is not straightforward to establish if the observed spatial variations
in OH distribution are due to diurnal or long-period variations. 
Most likely they are the result of a combination of factors such as 
the single and triple peaked light-curve 
variations with a period near 54--55 hr and the change in observing geometry 
between pre-CA and post-CA. 

Fig. \ref{fig:OHmaps} allows us to analyze the OH spatial distribution
closest to the nucleus, revealing structures at much smaller scales.
In both images the Sun is located on the right side while the different 
orientation of the nucleus is due to the 
movement of the spacecraft. The geometrical effect of the spacecraft motion is that images after 
CA are roughly upside down compared to those before the CA. A radial structure coming from the central waist of the nucleus 
is visible in both images, but more evident in Fig.~\ref{fig:OHmaps}b.

In order to measure reliably the opening angle and the angular extent 
of this feature we plot
the azimuthal profiles (Fig. \ref{fig:OHazprof}) 
of the OH column density averaged over rings of 1 km width, centered on the nucleus, 
respectively with radii of 5, 8, and 12 km.
The 
azimuth angle $\Theta$ is measured counterclockwise from the vertical line going from the center
of the nucleus perpendicularly to the bottom of the image.
The profiles have been normalized such that
the average value of each profile corresponds to the average of the profile
at 5 km, represented by the dashed line in the plots. The original average values
did not differ anyway by more than a factor 2.
Before CA (Fig.~\ref{fig:OHmaps}a) the OH jet has a peak around $\Theta\sim40^\circ$, 
while after CA (Fig.~\ref{fig:OHmaps}b)
it is directed towards $\Theta \sim 140^\circ$--150$^\circ$, 
which is compatible with the geometrical change between the two frames, 
indicating the persistence of the jet.
The OH radial structure has an opening angle of about 30$^\circ$
before CA which becomes broader after CA to about $40^\circ$ 
and extends up to about 12 km from the nucleus' surface. 
After CA it seems to have a second component blending down to the right of the 
central plume giving rise to a ``fountain-like''
structure. 

\begin{figure*}[t!]
  \gridline{
\fig{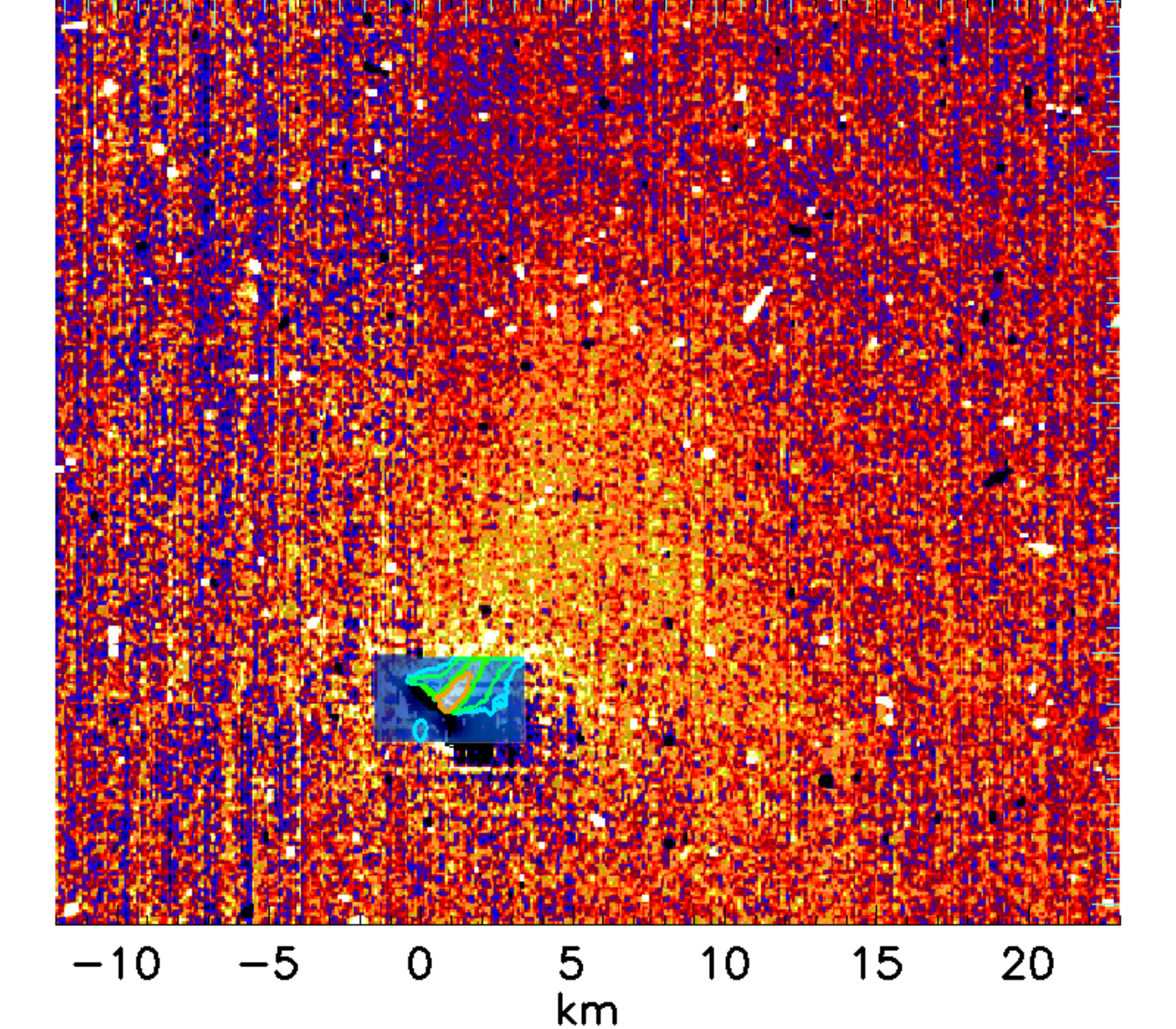}
{0.45\textwidth}{(a)}
\fig{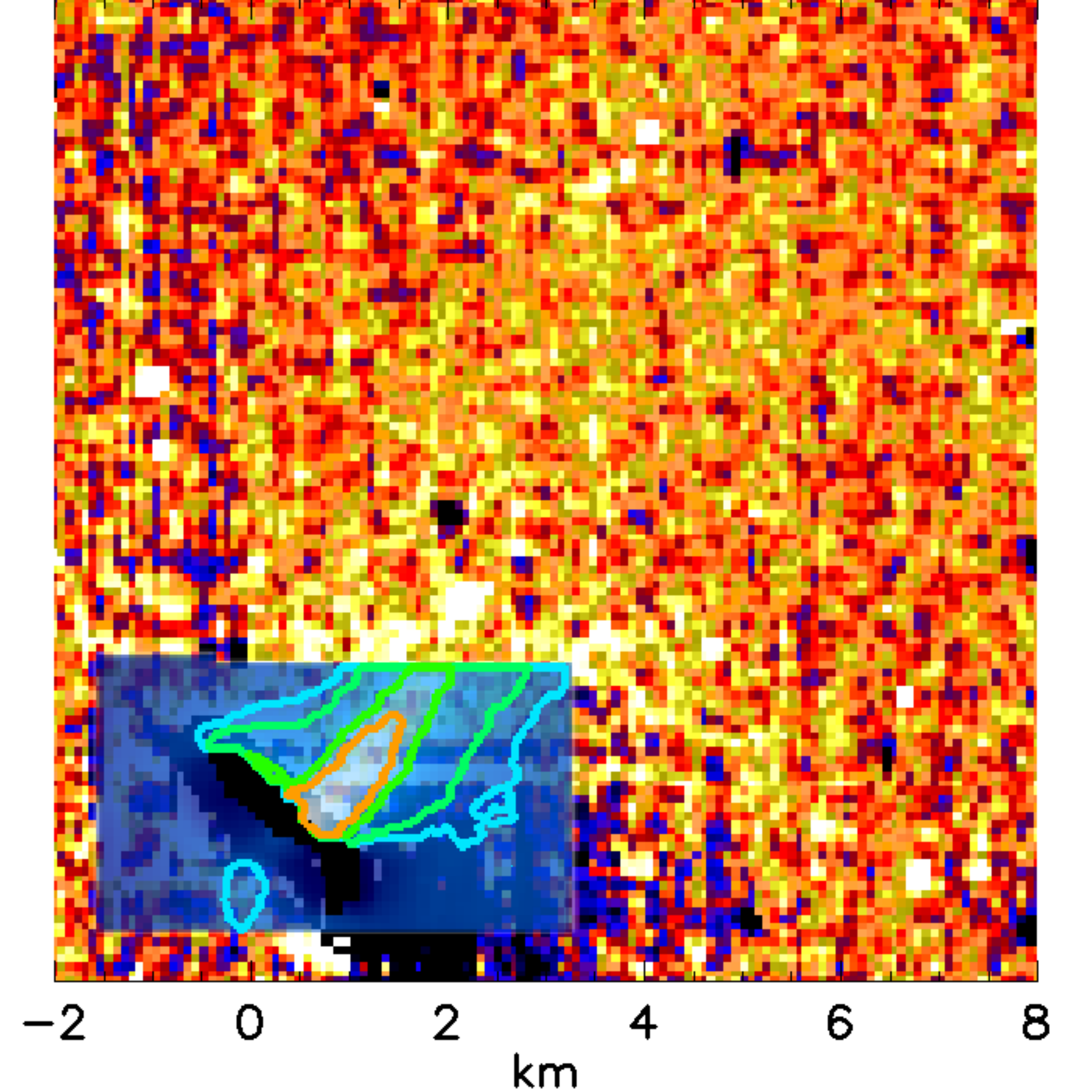}
{0.45\textwidth}{(b)}}
    \caption{(a) HRI-IR water vapor spectral map acquired 7 minutes
    after CA scaled, rotated and 
    shifted to be superimposed on 
    MRI-OH ID 5006065 column density map in Fig \ref{fig:OHmaps}b.
    (b) An enlargment of panel a within the near-nucleus region.}
    \label{fig:OHspmaps}
\end{figure*}  

 \begin{figure*}[hb!]
 \epsscale{1.01}
 \plotone{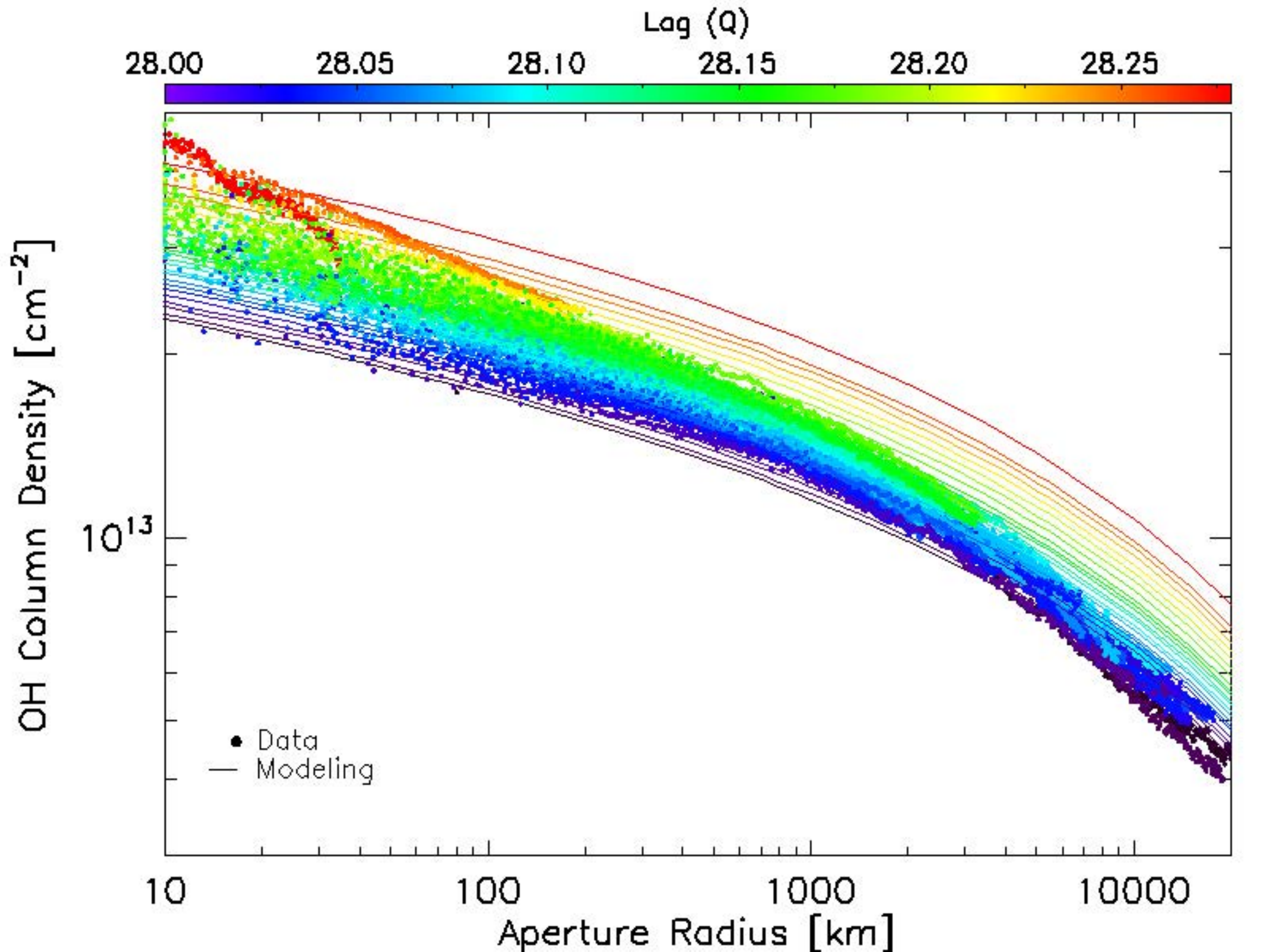}
   \caption{Radial profiles of OH column density azimuthally averaged for
   all 153 images acquired in the period DOY 301-311.
Solid lines show the Haser-based model profiles for various production rates values
according to the colorbar. The observed data points have the color of the best-fitting
model profile.}
  \label{fig:coldensprofall}
\end{figure*}
	
\begin{figure*}[!bt]
\gridline{
\fig{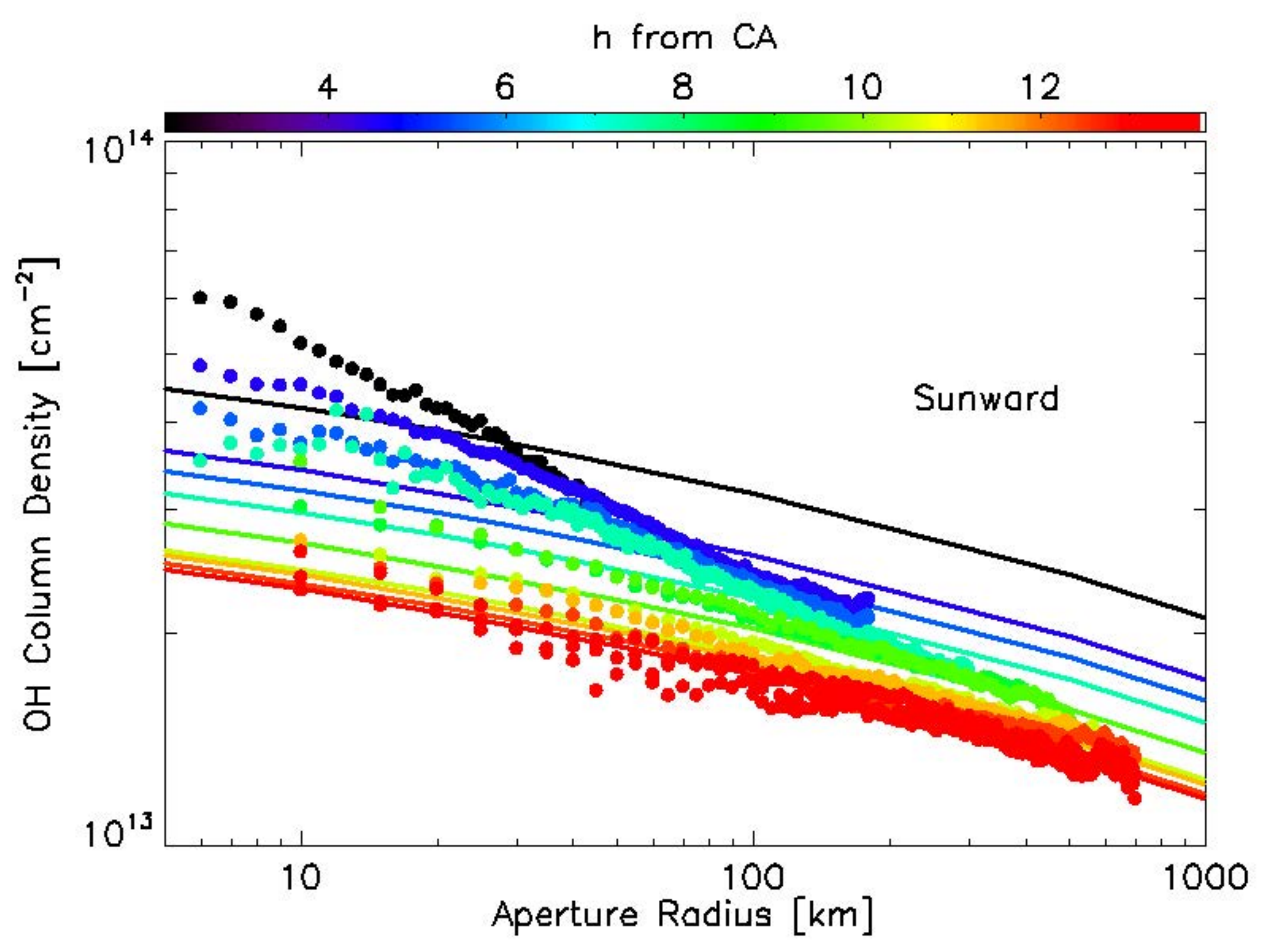}
{0.45\textwidth}{(a)}
\fig{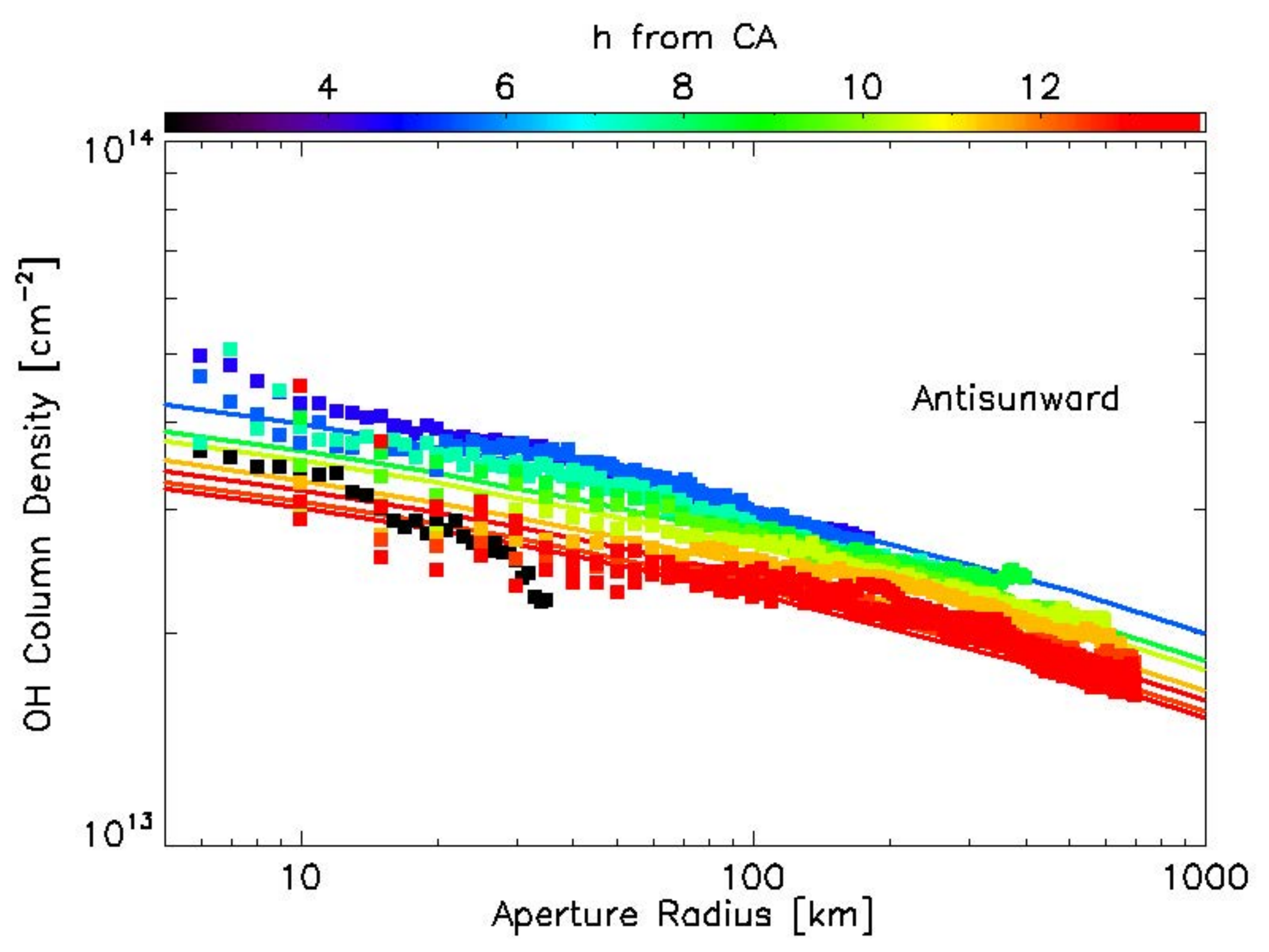}
{0.45\textwidth}{(b)}}
\caption{OH column density profiles sunward side (a) and 
anti-sunward (b)
   for a total of 12 images acquired between 10 min and 14 hours after CA.
   The colors of the data points indicate the acquisition time according to the colorbar.
   For each observed profile the best-fitting model profile has been plot
   with the same color of the observed profile. 
   The corresponding production rate values 
   are not indicated since the aim of the plot is to show the deviation
   of the observed profiles with respect to the models in the sunward direction 
   as opposed to the very good accordance
   obtained for the anti-sunward profiles.}
\label{fig:coldensprofdiv}
\end{figure*}

\setcitestyle{notesep={, }}
The observed spatial distribution differs from what we would expect
for fragment species. 
 Fluorescent OH transitions have lifetime of about 4000 s
at the heliocentric distance and velocity of Hartley 2 at the moment of 
observation. Such long lifetime would dimish the effects of potential spatial 
asymmetries in the distribution of the parent water molecule on the 
distribution of the fluorescent emission from OH.
However, the observed distribution resembles more a parent-like distribution.
We compared the OH structure visible in Fig.~\ref{fig:OHmaps}b with the water vapor spectral map
obtained on DOY 308 14:07 UTC (about 7 minutes after CA) by the HRI 
Infrared instrument (HRI-IR) on board DI \citep[Fig. 11A]{AHearn2011,Protopapa2014}.
The HRI-IR map has a spatial resolution of 55 m pix$^{-1}$ and covers a region of about 5 km, much smaller
than MRI FOV (70~km). It has been rescaled, rotated and shifted to the MRI image, 
so that the nucleus was superimposed in the two frames (Fig. \ref{fig:OHspmaps}).
The HRI-IR image shows a strong water vapor plume coming 
from the central waist of the nucleus. The OH structures in the MRI are clearly a continuation of the 
water vapor plume coming from the comet's waist seen by HRI-IR. 
 The strong spatial correlation between non-isotropic 
appearances of OH and H$_2$O suggests that a 
non-RFE component contributes to the total OH flux.
The excited OH* fragments, having a very short radiative lifetime 
of about 10$^{-6}$ s, instantly decay preserving the spatial distribution
of the parent molecule.

   
   \subsection{Radial Column Density Profiles}
       \setcitestyle{citesep={;}}
       
  \begin{figure*}[ht!]
  \epsscale{1.1}
   \plotone{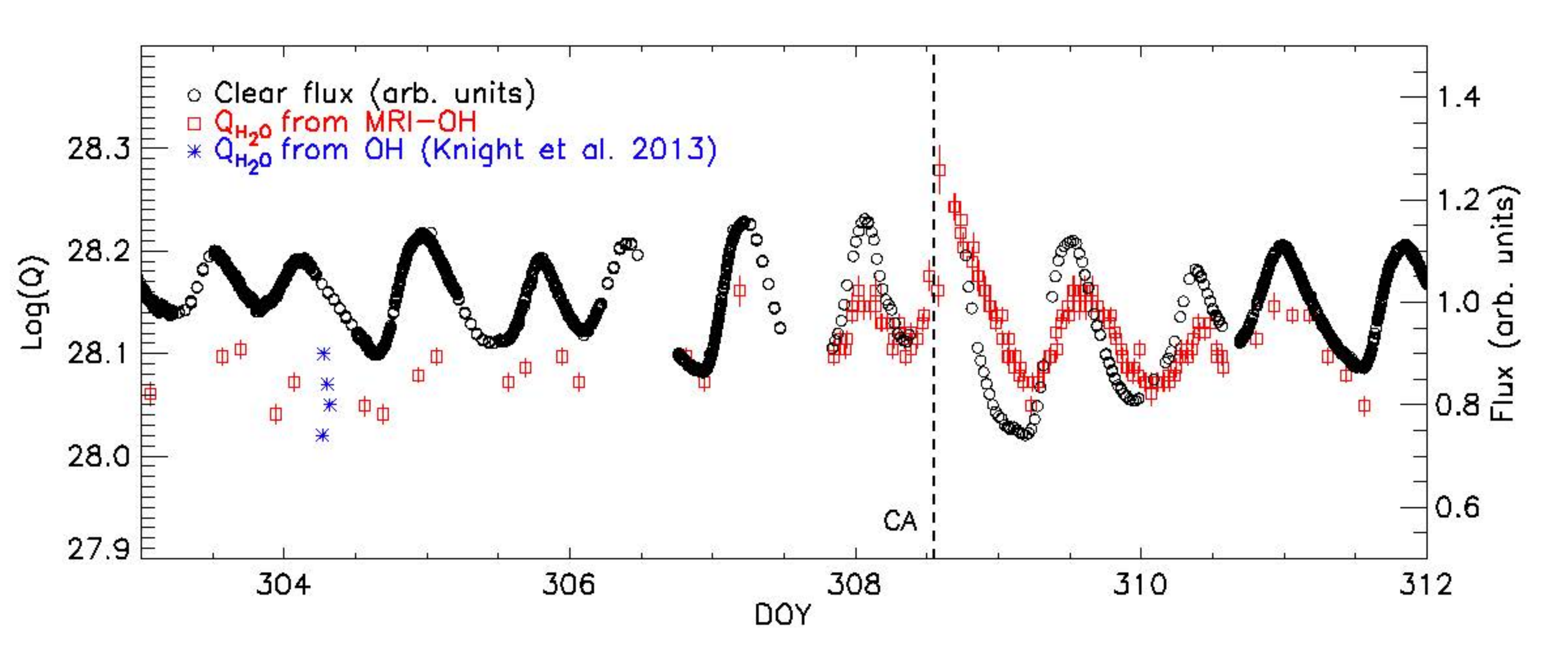}
   \caption{Temporal Variation of water production rate derived from MRI-OH observations
   using a two-generation Haser model (red squares), compared with Clear1 visible light curve (arb. scaled)
   within an aperture of 14 pixels (black circles) \citep{Williams2012}.
   Groundbased water production rates by \citet{Knight2013} are also shown for 
   comparison.}
  \label{fig:LC1}
\end{figure*}
       
\noindent For each image of the dataset we extracted an azimuthally 
averaged radial column density profile. 
The full dataset provides a good coverage of the coma 
between 10~km and a few thousand kilometers from the surface.
 Given the anti-sunward enhancement observed in the OH spatial 
 distribution 
pre- and post-CA (Fig.~\ref{fig:CDmaps_pre} and \ref{fig:CDmaps_post}) 
and the strong sunward radial jet observed at CA (Fig. \ref{fig:OHmaps}),
we separately studied sunward and anti-sunward profiles inside
1000 km from the nucleus in 12 images acquired between 
10 minutes and 14 hours after CA.
 
 The 153 OH column density profiles are shown in 
Fig.~\ref{fig:coldensprofall}. They are overall consistent and form a regular
trend without significant
discontinuity. 
The sunward profiles (Fig. \ref{fig:coldensprofdiv}a), acquired between 0 and 14 hours from CA, show a rapid change both in absolute value and in their shape. 
 The observed profiles become increasingly steeper approaching 
  CA and the OH
  column density at 10~km from the nucleus doubles in about eight hours compared to the column density before CA.
  The anti-sunward profiles (Fig. \ref{fig:coldensprofdiv}b)
  remain flatter than the sunward profiles.

We compare the azimuthally averaged column density profiles 
with modeled profiles. 
The \citet{Haser1957} model describes the distribution of fragment species. 
It assumes a 
spherically symmetric source of uniformly outflowing parent molecules 
that photodissociate with an exponential lifetime to create fragment species. 
In the vectorial model \citep{Festou1981} 
the daughter molecule has an excess velocity from the photodissociation of the parent.
As a consequence the daughter does not have the same outflow velocity of the parent but may have an 
isotropic distribution around the parent molecule. 

\noindent Despite the limitations of the standard Haser model \citep[{see, e.g., }][]{Crifo2004}
it is a commonly used model for a direct comparison of different observations owing to its simplicity.
Moreover, in the inner coma collisions enforce a radially outward flow
\citep{Woodney2002} thus making the Haser model still useful for spatial 
radial profile fits in these regions.
Since a more detailed and complex numerical model, 
such as the Monte Carlo and the updated DSMC calculations \citep[{see, e.g., }][]{Combi1988,Tenishev2008},
is out of the scope of this paper and a simple relative comparison is sufficient here, we decided to 
proceed with a modified two-generation Haser model.
We used the transformation 
equations in \citet{Combi2004}
that relate a realistic spatial profile obtained with the vectorial model
to the equivalent Haser scalelengths and lifetime. We adopted the typical values for 
gas velocity of $v=1$ km s$^{-1}$, lifetime of water 
$\tau_{H_2O}=0.96\times 10^5$ s 
and lifetime of OH $\tau_{OH}=1.5\times 10^5$ s 
\citep{Combi2004}.
We ran computations 
for a series of production rates ranging between $10^{28}$~mol~s$^{-1}$ 
and $2\times 10^{28}$~mol~s$^{-1}$ with incremental steps of $0.02-0.05\times10^{28}$~mol~s$^{-1}$.
The modeled number densities are then integrated along the line of sight and 
used in comparison with the observed column densities profiles. 
The model with the smallest residuals represents the best fit
among the computed curves.
 The general agreement found between 
the observed and the modeled profiles -- except
for the very inner coma -- justifies our approach.
 
  The modeled column density profiles are shown as solid lines in 
Fig. \ref{fig:coldensprofall} and \ref{fig:coldensprofdiv}.
They adequately fit most of the column density profiles apart from the 
very innermost regions (inside 20 km) observed at close distances
in the sunward direction, 
where the data show higher and steeper 
column densities with respect to the model.
 The measured antisunward OH profiles show an increase in the OH column density as we get close to the CA
  that is much more in accordance with the modeled column density 
   profiles. A possible explanation for this dichotomy is the presence of two OH
    emission components: the first a tailward component, stable in time and 
   changing with the rotation 
  of the nucleus, and compatible with the predicted fragment species distribution of OH; 
  and a second component, mainly sunward, deviating from the model prediction and 
  requiring a much higher number of OH molecules, and likely correlated with the water gas plume
  observed in HRI-IR maps \citep{AHearn2011}. 
  The OH column density in the sunward direction can be
  due to a different emission process in addition to fluorescence. 
  We will discuss this further in the next section.

\subsection{Production Rates}

 The adopted model is based on the assumption that OH is produced solely 
 by photodissociation of H$_2$O, 
 and that OH emission is driven by RFE.
 The best fit of each column density profile is then 
 used, under these circumstances, to derive the water production rate.
 
   The temporal variation of the water production rate is shown with red empty squares in Fig.~\ref{fig:LC1}.
   We compare this with the visible light curve 
   observed through the Clear1 filter within 14 pixels aperture 
   (black filled circles) \citep{Williams2012}
   to investigate the correlation between gas and dust activity.
   Ground-based water production rates by \citet{Knight2013} 
   are also overplotted for comparison.
   The agreement between the dust and gas light curves, in particular outside the CA peak, 
   suggests that the OH emission correlates well with the 
   dust content of the coma and the rotation of the comet,
   despite the
   different escape speed of the two components. The latter is likely 
   responsible for the small shift between the curves.
   The amplitude of the water production rate is smaller
   than the amplitude of dust light curve 
   in this relative scale, except for the peak at CA, which shows 
   a much higher water production rate with respect to all other peaks. 
   Since no physical relevant changes are expected in the comet's activity 
   corresponding to the approach of the spacecraft, 
   apart for an increased spatial resolution, the increased amplitude 
   of the CA peak is further evidence that a mechanism 
   other than fluorescence drives part of the OH emission.

   \begin{figure}[h!]
   \epsscale{1.}
   \plotone{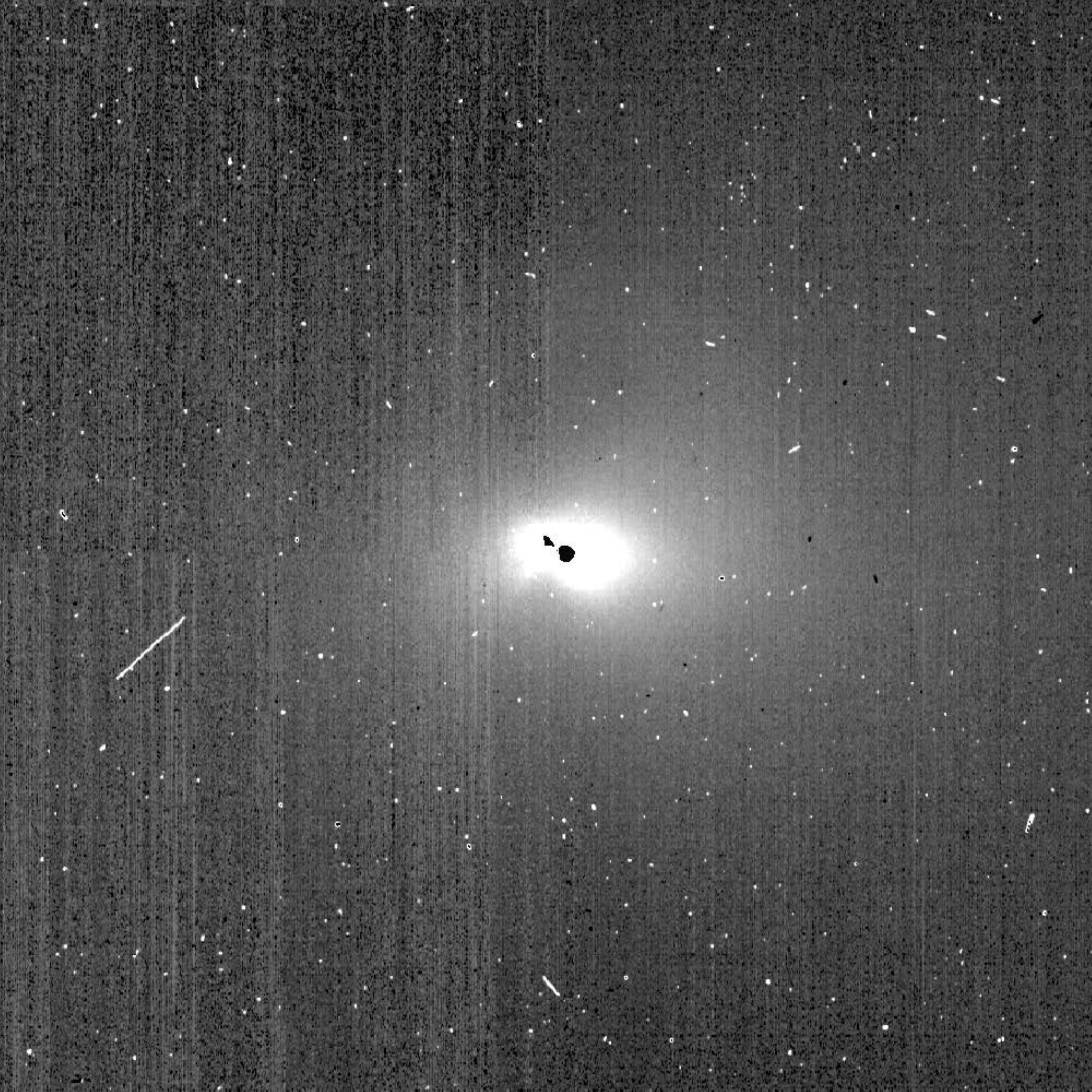}
   \caption{Resulting continuum image subtracted 
   from OH image ID 5006065. 
   A linear stretch of the square root histogram of the image is used
   for visualization.}
  \label{fig:continuo}
\end{figure}

\section{Discussion}
\label{disc}
\subsection{Emission Processes}
        \setcitestyle{citesep={;}}
        
        There are two clear indications that OH fluorescence 
        is not the sole mechanism producing the observed emissions
        through the MRI-OH filter. 
         First, the data show a strong jet in the innermost coma of Hartley 2, 
         extending up to 12 km from the nucleus
         (see Fig. \ref{fig:OHmaps}), with a significant 
         correlation with the water gas distribution 
         observed through HRI-IR (see Fig. \ref{fig:OHspmaps}). 
         Second, the column density profiles deviate from an Haser-based model
         only in the very vicinity of the nucleus in the sunward
         direction, resulting in an excess peak in the derived water production rate at CA that is not expected from an increase in spatial 
         resolution only (see Fig. \ref{fig:coldensprofdiv}a).
         
             The jet visible in OH at CA is not present in the continuum image
     (Fig. \ref{fig:continuo}), which 
     has a completely different morphology with a more circular 
     distribution around the two lobes of the comet, 
     in agreement with the dust distribution 
     observed by HRI-IR and HRI-VIS
     \citep{Protopapa2014}.
     We therefore rule out that the OH feature can be attributed to a residual dust jet.
              
      We investigated the possibility that this emission is produced by a species
      other than OH. The only feature at these 
      wavelength which may contribute to the observed emission is 
      the B$^3\it{\Sigma}_u-$X$^3\it{\Sigma}_g$ band of S$_2$ molecule.
      S$_2$ has a short lifetime of a few hundreds seconds at 1 au,
      so it is present only very close to the nucleus.
      
      The S$_2$ fluorescence rates are approximately 4$\times10^{-4}$ ph s$^{-1}$ mol$^{-1}$ \citep{Ahearn1983}, comparable to the OH fluorescence efficiency at Hartley 2 
        ($2.4 \times 10^{-4}$ ph s$^{-1}$ mol$^{-1}$). 
        S$_2$ has been observed in a small number of comets and abundances 
        with respect to water are of order 0.001--0.1\% 
        \citep{Ahearn1983,dealmeida1986,Kim1990,Weaver1996,Laffont1998,Bockelee-Morvan2004}. 
        Because the number of OH molecules is at least 100 times larger than the number of S$_2$ molecules we expect its contribution to the observed emission to be very small.

     Another possibility is that excess OH comes either directly from
     the nucleus, or from slowly moving grains in the
coma, in a similar way that CN seems to have these contributions \citep{Bockelee-Morvan1985,Fray2005}.
If the ``fountain-like'' structure observed in the post-CA 
(Fig. \ref{fig:OHmaps}b) is real, 
its presence would support the hypothesis that OH is coming from 
slowly moving grains that
are fountaining out over a significant portion of the rotation arc.
However, this scenario would not explain the very good correlation 
observed in the inner coma between OH and water gas (Fig. \ref{fig:OHspmaps}).
     The direct emission from excited OH* molecules is instead
     a plausible candidate to explain our observations.
     The two most likely processes to produce excited OH* from H$_2$O 
     in the coma are dissociative electron impact excitation 
     (DEIE) and emissive photodissociation, also called prompt emission (PE),
     respectively:
     \begin{eqnarray}
     \textnormal{DEIE}:&\hspace{0.5cm} H_2O + e \to OH^* + H \\
    \textnormal{PE}:& \hspace{0.5cm} H_2O + h\nu \to OH^* + H
    \end{eqnarray}
The production of excited water group fragments (O*, OH*, OH*$^+$) by electron impact has been 
observed in the inner coma of comet 67P/Churyumov-Gerasimenko by the {\it Alice} and OSIRIS 
instruments on board the Rosetta spacecraft \citep{Feldman2015,Bodewits2016}.
Like MRI, the OSIRIS instrument was equipped with narrowband filters to image the gas and dust in the coma.
Similar to the results described here, it observed an excess of emission in the OH filter 
(and in its OI, NH, and CN filters), and the morphology of the emission
indicated a process that directly produced OH in the $A^2\it{\Sigma}^+$ state. 
The surface brightness in the inner 100 km of the coma was orders of magnitude
larger than could be explained by photo processes. As the comet got closer to the Sun 
and its production rates exceeded several time 10$^{27}$ molecules s$^{-1}$, 
emission levels dropped to 'normal' levels. \citet{Bodewits2016}
attributed this to decreasing electron 
temperatures; below 10~eV electron impact collisions lead to ro-vibrational excitation of 
H$_2$O molecules rather than to dissociative or ionizing reactions. 

Both DEIE and PE produce OH emission that falls within the passband of 
MRI-OH filter (next section) and the Deep Impact spacecraft was not equipped with any plasma 
instruments that would allow us to distinguish between the two 
emission processes. However, the EPOXI observations of 103P/Hartley 
2 were acquired at 1.064~au, when the comet had a water production 
rate of 10$^{28}$ molecules s$^{-1}$. The physical environment in the inner 
coma then was comparable to that of 67P/Churyumov-Gerasimenko during 
perihelion when photo 
processes rather than electron impact drove the emission. 
That environment is likely also analogue to the inner coma of comet 
Hyakutake, for which \citet{AHearn2015} used high-resolution spectra to distinguish PE from DEIE and PE. 
We therefore find the prompt emission scenario more likely than the electron impact scenario and
 will quantitatively test this hypothesis in the next section.

\subsection{Prompt Emission Scenario}
     
The fluorescence mechanism mainly populates the first 5 rotational levels of OH*$(A^2\it{\Sigma}^+$). 
 \citep{Schleicher_AHearn1988}.
The direct population of OH*($A^2\it{\Sigma}^+$) by photodissociation 
of water has been studied by many authors. \citet{Carrington1964}
found that the rotational levels from 18 to 23 are the most populated 
states, supported by later studies by \citet{Yamashita1975} and 
\citet{Mordaunt1994}. \citet{Harich2000} also found that 
highly rotationally excited OH(A) products are dominant. 
This result has been also recently confirmed 
by the observation of OH PE lines in the spectrum of comet Hyakutake
\citep{AHearn2015}.

\begin{figure}
 \epsscale{1.2}
 \plotone{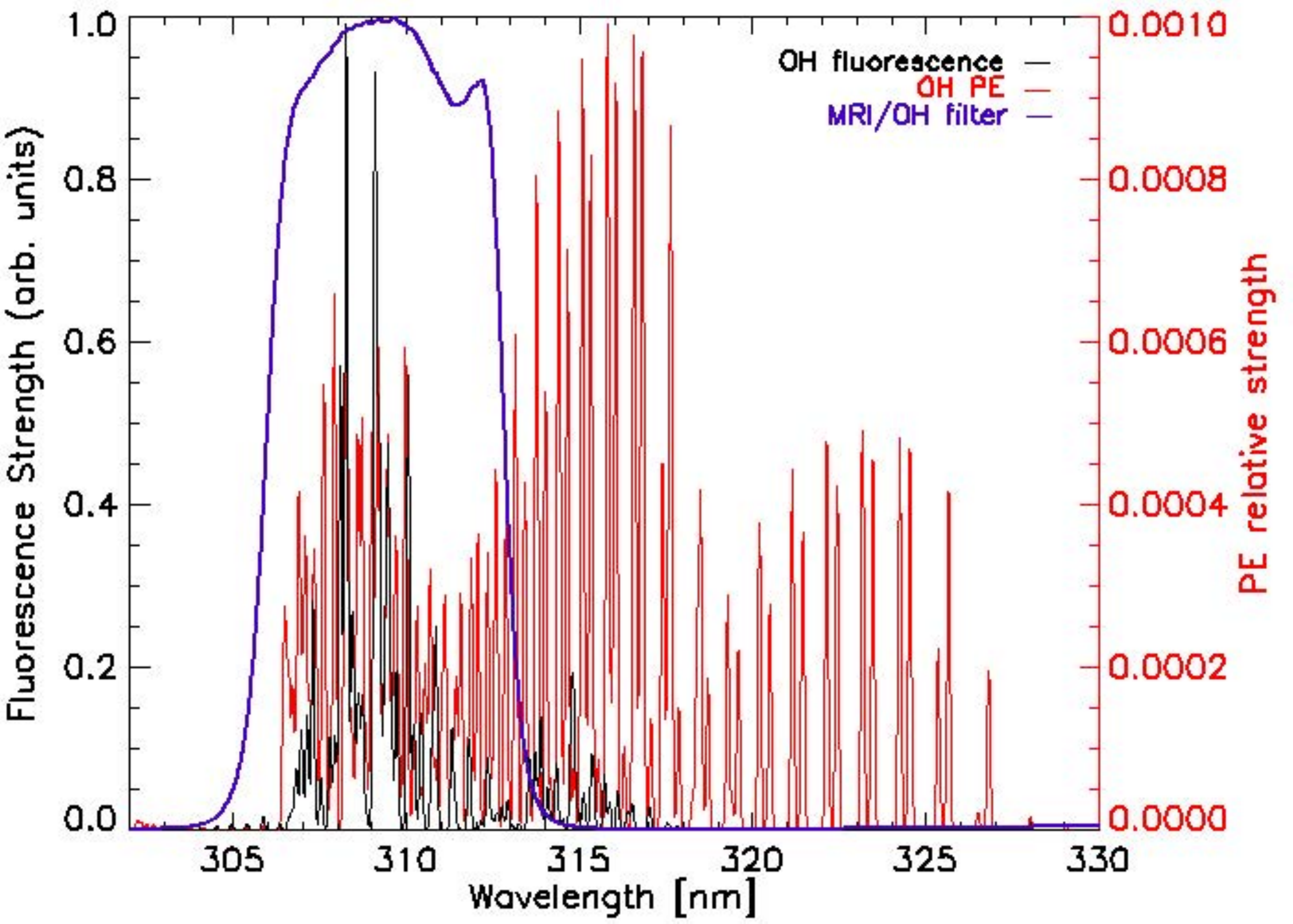}
 \caption{OH Fluorescence (black) and prompt (red) emission 
  spectral distributions obtained with LIFBASE software 
  and population distribution by 
  Schleicher (priv. comm.) and \citet{Carrington1964} respectively.
  Violet curve shows the MRI-OH bandpass filter weighted
   for the CCD QE and mirror reflectivity.}
   \label{fig.fl_pe_spectrum}
\end{figure}

The excited OH*$(A^2\it{\Sigma}^+$) molecules directly decay emitting 
at slightly higher wavelengths than fluorescence and with a different intensity in the NUV.
We produced synthetic OH spectra for both 
prompt emission and fluorescence emission using the software 
LIFBASE \citep{Luque1999}, assuming population distributions 
for PE from \citet{Carrington1964} and Schleicher's (priv. comm.) for RFE. 
The resulting synthetic spectra 
are shown in Fig. \ref{fig.fl_pe_spectrum} together with the MRI-OH bandpass.
The PE spectrum is broader and centered at higher wavelength,
however a non negligible fraction, $f_{PE}$, passes through the MRI-OH filter. 
Using Eq.~\ref{eq:convolution} we calculated this fraction
by weighting the PE spectrum with the MRI-OH 
filter bandpass, mirror reflectivity and CCD quantum efficiency, 
and find that a fraction of $f_{PE}=38.75\%$ of all PE emission is sampled by the MRI and its OH narrowband filter.
For comparison, the Rosetta/OSIRIS OH filter \citep{Keller2007} has
a slightly narrower bandpass (4 nm) and blocks longer wavelengths than the MRI filter, 
making it less sensitive to OH PE.

The theoretical contribution $S_{PE}/S_{FE}$ of the two mechanisms, as function of 
the projected distance $\rho$ from the comet's nucleus in the sky plane, 
can be calculated by:

\begin{equation}
 \frac{S_{PE}}{S_{FE}}(\rho)=\frac{N_{\text{H}_2\text{O}}(\rho)}{N_{\text{OH}}(\rho)} 
 \frac{D Br(\text{OH}^*(A^2\it{\Sigma}^+)) f_{PE}}{ g'_{\text{OH}}}
  \label{eq.relative_strength_filter}
\end{equation}

\noindent where $N_{\text{H}_2\text{O}}(\rho)$ and $N_{\text{OH}}(\rho)$ are the column densities of H$_2$O molecules and OH molecules at the 
projected distance from the nucleus $\rho$; $D$ is the photodestruction rate of water molecules, i.e. the total number of water 
molecules which actually photodissociate per second; $Br(\text{OH}^*(A^2\it{\Sigma}^+))$ is the branching ratio 
for the production of OH*$(A^2\it{\Sigma}^+)$, i.e. the fraction of OH molecules left in the excited $A^2\it{\Sigma}^+$ electronic state
relative to the total number of OH radicals produced by photodissociation; $g'_\text{OH}$ 
is the effective fluorescence efficiency described in Sec. \ref{sec:OHfluorescence}, 
and $f_{PE}$ is the fraction of the PE spectrum visible through the MRI-OH filter.

 \begin{deluxetable}{cccc}
 \tablenum{2}
 \tablecaption{Lifetime, photodestruction rate of H$_2$O and branching ratios for OH*($A^2\it{\Sigma}^+$)
  \citep{Combi2004} \label{tab.parameters}}
  \tablewidth{0pt}
  \tablehead{
\colhead{} & \colhead{$\tau_{\text{H}_2\text{O}}$} 
& \colhead{$D$} & \colhead{$Br(\text{OH}^*)$} \\
\colhead{} & \colhead{[s]} & \colhead{[mol s$^{-1}$]} & \colhead{[$\%$]}}
  \startdata
  Quiet Sun & 9.6E4 & 1.04E-5 & 3.6 \\
  Active Sun & 7.1E4 & 1.41E-5 & 4.1 \\
  \enddata
 \end{deluxetable}

The ratio $\frac{N_{\text{H}_2\text{O}}}{N_\text{OH}}(\rho)$ in Eq. \ref{eq.relative_strength_filter} 
has been estimated using the two-generation Haser model adopted in the 
previous section. Given the low activity of the Sun at the time
of the encounter we used the quiet Sun values 
from \citet{Combi2004} summarized in Tab.~\ref{tab.parameters}. 
The ratio $\frac{N_{\text{H}_2\text{O}}}{N_{\text{OH}}}(\rho)$ is independent of the water 
production rate value and decreases exponentially with the projected 
distance from the nucleus. The photodestruction rate and the PE branching ratio depend mainly 
on the solar activity.   
 
\begin{figure}[ht!]
\epsscale{1.2}
 \plotone{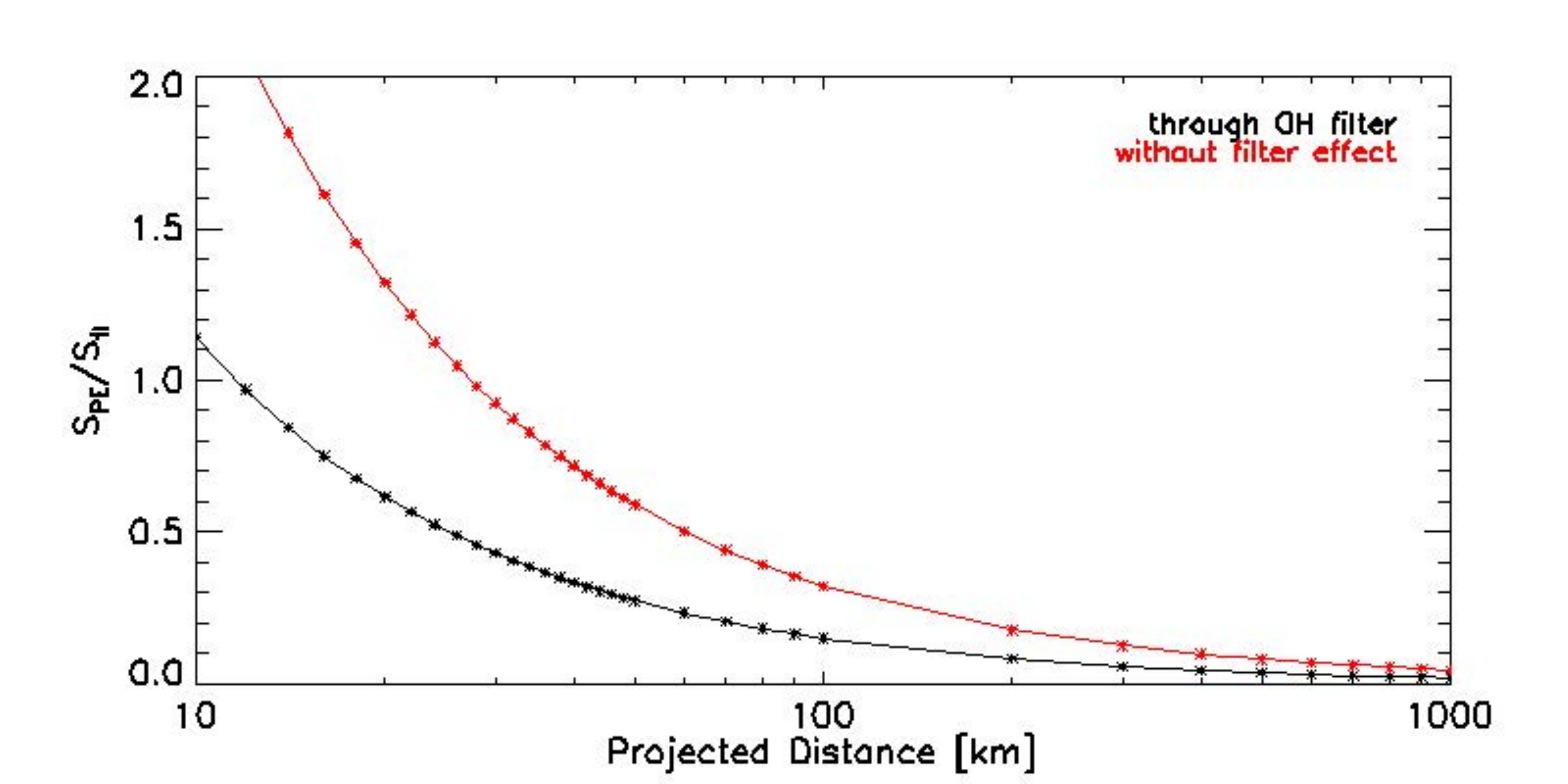}
 \caption{Relative strength of OH prompt emission to fluorescence emission
 as function of the projected distance from the nucleus 
modeled using a two-generation Haser model, as seen through MRI-OH 
filter (black line) and without the effect of the filter (red line). }
  \label{fig.tot_pe_fluor}
\end{figure}
 
The total theoretical ratio $S_{PE}/S_{FE}$ in Eq.~\ref{eq.relative_strength_filter} 
as seen through the MRI OH narrowband filter, and the same ratio without 
the effect of the filter, have been computed for $\rho$ 
between 10 and 1000~km and are shown in
Fig. \ref{fig.tot_pe_fluor}.
Close to the nucleus the PE is as efficient as RFE
but outside $\sim$20 km fluorescence becomes the dominant emission 
mechanism. This analysis suggests that a non negligible fraction 
of the OH flux in the innermost coma 
may be actually attributed to OH prompt emission.

\begin{figure}[ht!]
\epsscale{1.2}
\plotone
 {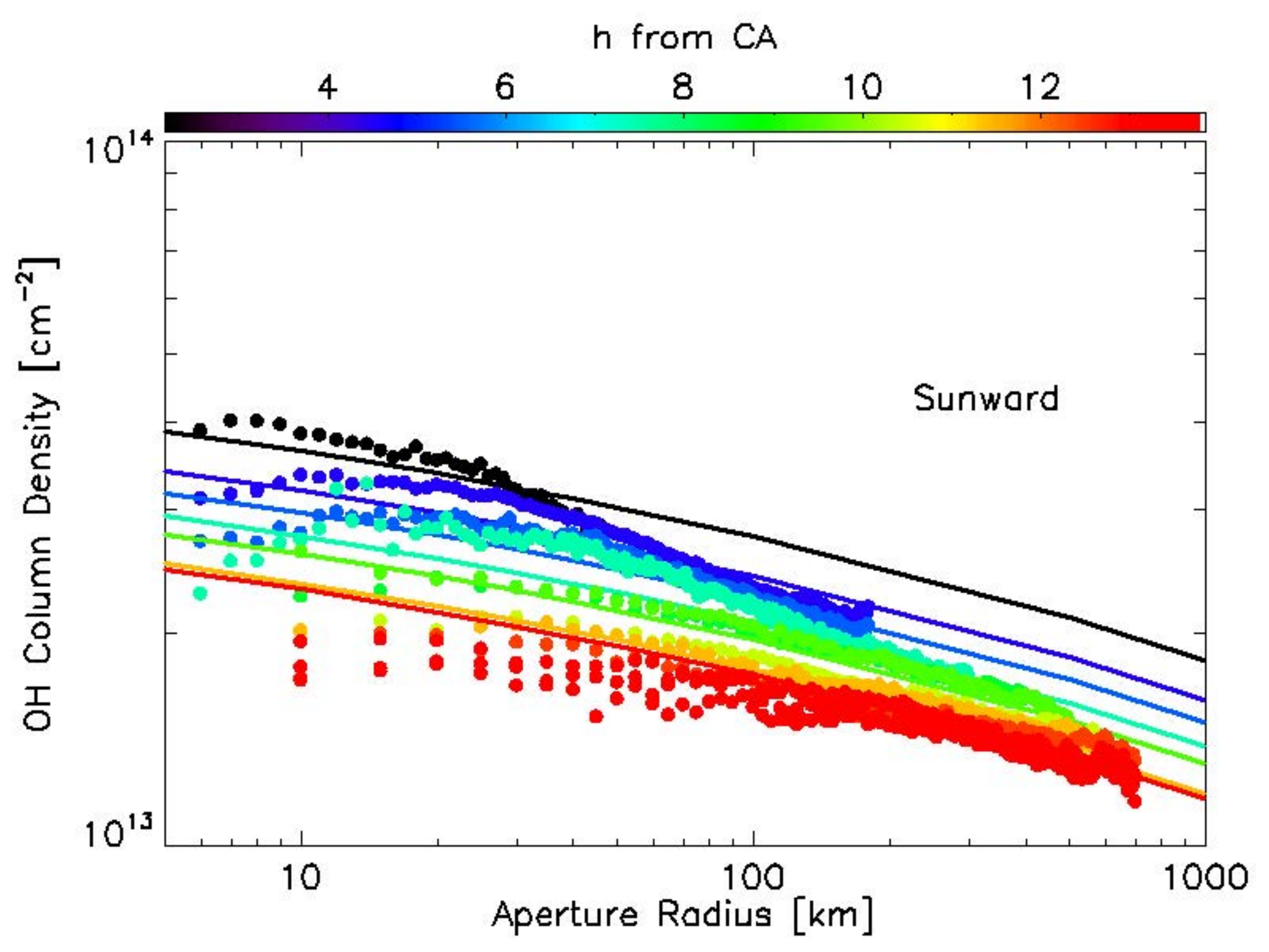}
 \caption{ Sunward column densities profiles 
 as in Fig. \ref{fig:coldensprofdiv} 
 but corrected for 40\% of the theoretical estimated prompt emission.}
  \label{fig.prodratePE0.4}
\end{figure}

    If the dominant emission mechanism is indeed PE,
    we would overestimate the OH column densities derived assuming fluorescence efficiencies.
    To retrieve pure fluorescence column densities, we used the total strength ratio $S_{PE}/S_{FE}$ 
    to subtract the expected PE fraction from the measured OH 
    column density.
The total theoretic fraction results higher than the value which is needed 
for the profiles to agree with the models.
If we subtract 40\% of the theoretical PE computed, 
 the modified column density profiles and the models agree much better (Fig. \ref{fig.prodratePE0.4}), 
even in the innermost coma. The models are still not an adequate fit to
the residual pure fluorescence column density profiles, 
but this can be attributed to the limitations of the 
Haser-based model for the inner coma.

\begin{figure*}[ht!]
\epsscale{1.1}
\plotone
 {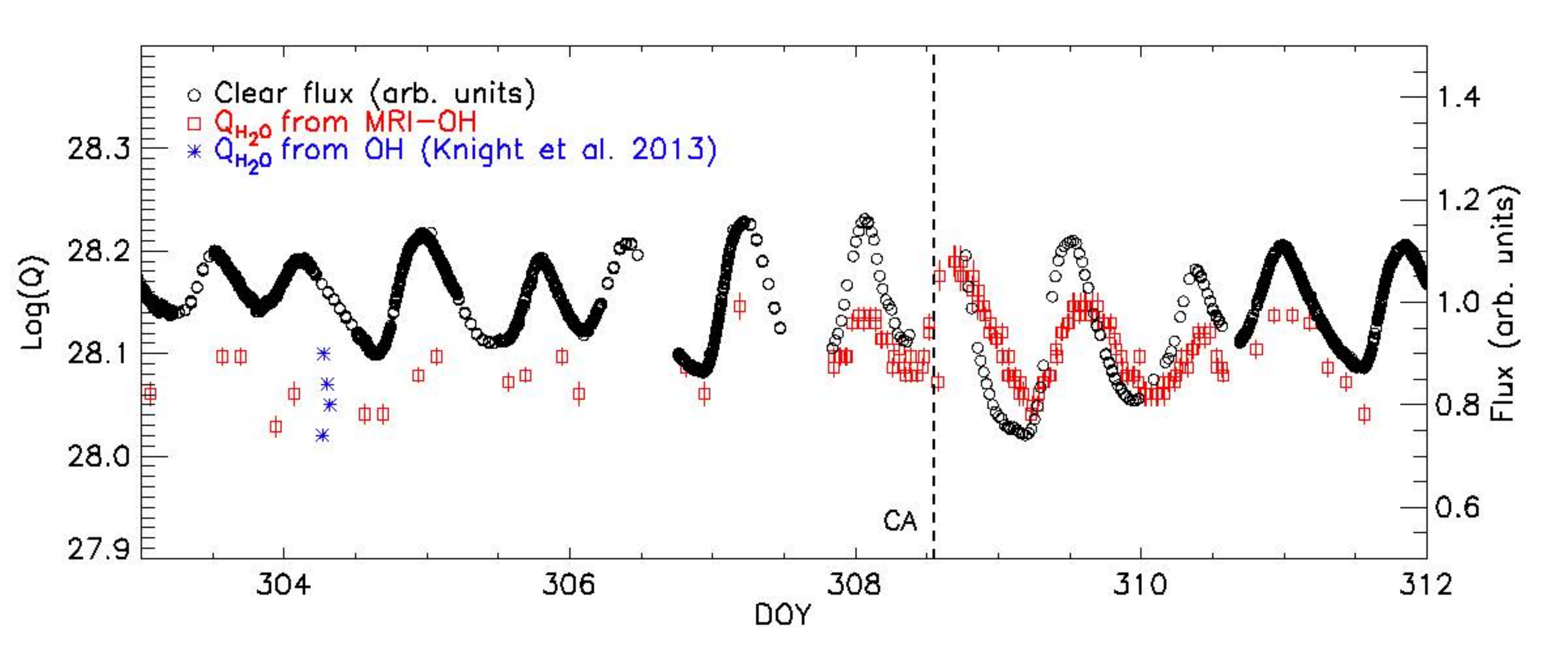}
 \caption{ Water production rate (red squares) corrected for 40\% of the theoretical estimated 
 prompt emission compared with Clear filter light curve (arb. scaled) within 
 aperture of 14 pixels (black circles) \citep{Williams2012}. Ground based water production rates
 by \citet{Knight2013} are plotted for comparison.}
  \label{fig.prodratevarPE0.4}
\end{figure*}

The water production rate variation curve has been also corrected assuming 
the same adjusted prompt emission factor
 (Fig. \ref{fig.prodratevarPE0.4}).
The strong CA peak is remarkably reduced if compared with Fig. \ref{fig:LC1} and is now more in line 
with the periodicity variation.

    This suggests that the PE mechanism is able to explain the data
within a factor two, which is acceptable given the large uncertainties in the 
numerical parameters used in the modeling and considering the several 
physical processes that we did not include in the calculations.
Photodestruction rates and branching ratios may vary significantly due
to the solar activity and they are affected by large uncertainties 
in calculated reaction cross sections. 
The assumed velocity and lifetimes of the fragments 
might vary by several factors in this very complex region of the coma
with respect to the typical assumed values.
The fraction $f_{PE}$ 
strongly depends on the assumed rotational and vibrational population distribution, 
which has its intrinsic uncertainties (see for example \citet{AHearn2015}).
The optical depth effect is not included in the calculations 
because the PE lines are optically thin but
the inner coma is opaque to solar Ly$_\alpha$. 
Moreover, collisions in the innermost coma 
could in principle de-excite a fraction of OH* molecules, thus reducing the effective PE observed. 
Given the very short lifetime (10$^{-6}$ s) of these fragments 
\citep{BeckerHaaks1973,AHearn2015}, very high molecular densities
would be required for collisions to significantly de-excite the OH* radicals.
However, given the importance of electron impacts observed by
Rosetta at comet 67P/Churyumov-Gerasimenko \citep{Feldman2015,Bodewits2016}, we carried out a rough estimate of the collisional rate of OH* molecules 
with electrons and neutrals (mainly water molecules) in the inner coma. 

\subsection{Collisional Quenching}

The collisional rate, $R_c$ is given by:

\begin{equation}
 R_c=\sigma_cn(\rho)v
\end{equation}
where $\sigma_c$ is the collisional cross section, $n(\rho)$ is the density of the collisional partner 
at cometocentric distance $\rho$, and $v$ is the typical relative speed. 
Collisions with electrons are more efficient in de-exciting the 
OH radicals than collisions with neutrals because the 
collisional cross section of electrons and OH* is 
about $10^{-10}$~cm$^{2}$ \citep{Schleicher_AHearn1988}, while the total
cross section of impacts by neutrals is about $1-5 \times 10^{-14}$~cm$^{2}$ 
\citep{Bockelee-Morvan2004}.

The Langmuir Probe (LAP) and Mutual Impedance Probe
(MIP) on Rosetta measured the electron number density 
at 10 km from the comet 67P/Churyumov-Gerasimenko when it was at 2.3 AU
from the Sun \citep{Edberg2015}. If we assume that
the number density scales linearly with the water production rate and the
photoionization rate, and thus the square of the heliocentric
distance \citep{Bodewits2016}, the electron number density at Hartley 2
at about 50 km from the nucleus would be about 10$^4$ cm$^{-3}$.
\citet{Bhardwaj1996} developed models of electron density profiles 
based on observations of comet Halley, whose production rate was 5-10 times
higher than Hartley 2, finding electron density of 10$^5$ cm$^{-3}$ 
at 50 km from the nucleus.
\citep[see also][]{Biver1999}.
 We conservatively assumed as upper limit the electron density 
profile modeled by \cite{Bhardwaj1996} for higher activity conditions 
to calculate the collisional rate.

The number density
of water molecules, certainly the most abundant species in the inner coma, 
is about $3.7 \times 10^8$~cm$^{-3}$, 
calculated using the two-generation Haser model assuming a production 
rate of $1.17 \times 10^{28}$ mol s$^{-1}$, the average of the water production rate
derived from MRI-OH observations.

\begin{figure}
\epsscale{1.2}
\plotone
 {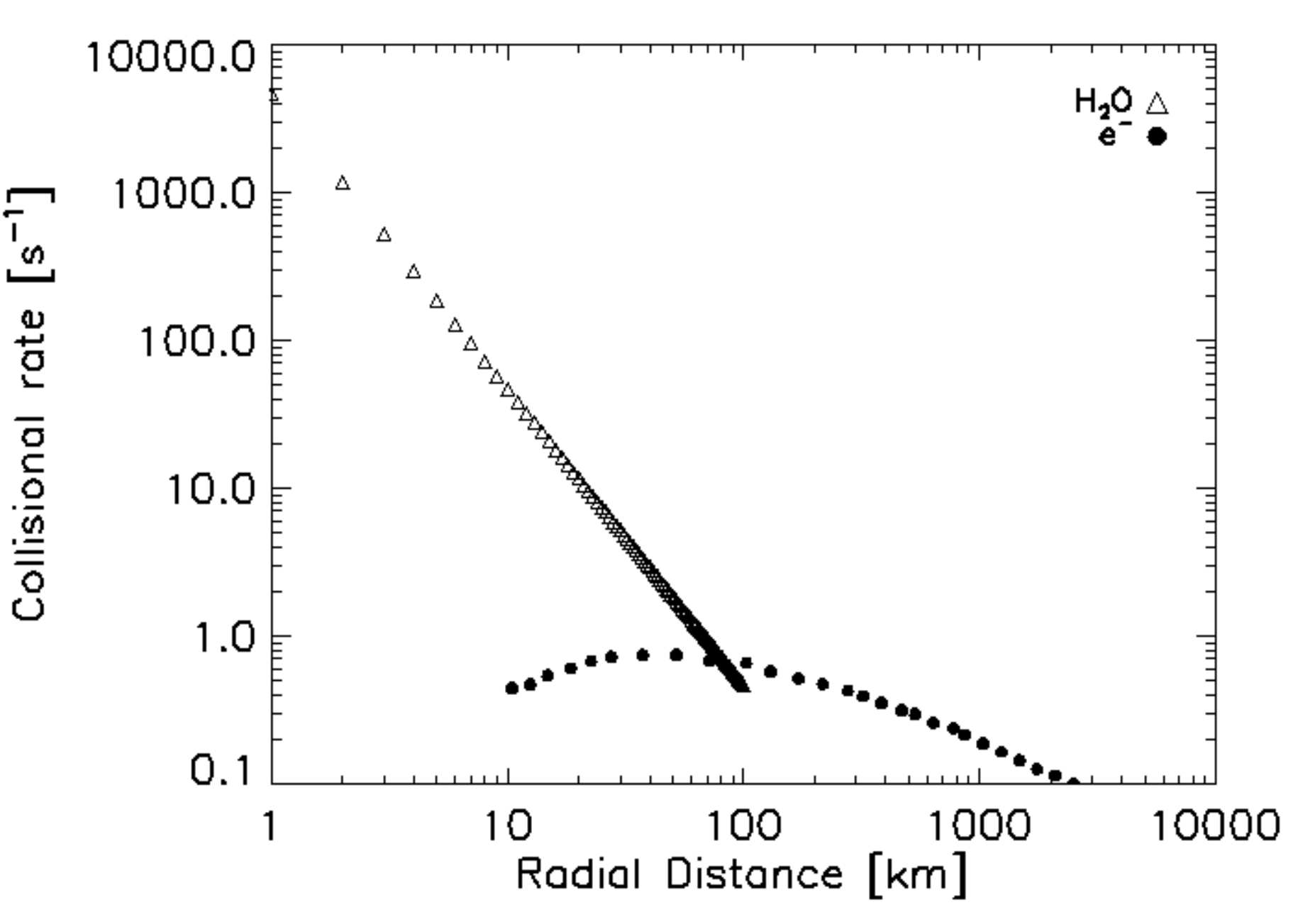}
 \caption{Collision rate of OH* molecules with 
 electrons (filled circles) and water molecules
 (open triangles) as function of the projected distance from the comet
 based on \citet{Bhardwaj1996} electron density and 
 water molecules density derived using an Haser-based model 
 and a production rate of $1.17 \times 10^{28}$ mol s$^{-1}$, the average of the water production rate
derived from MRI-OH observations.}
  \label{fig:collrate}
\end{figure}

The collisional rates of OH radicals with both electrons,
and water molecules are shown in Fig. \ref{fig:collrate} as a function 
of the radial distance from the comet. 
 Even when we assume the upper limit for the electron density 
we find that the collisions with neutrals are dominant 
in the innermost regions up to about 100 km since the local ratio 
n$_{\text{H}_2\text{O}}$/n$_{e^-}$ is very large \citep[see also][]{Bockelee-Morvan2004}.

At 10 km from the nucleus the collisional lifetime $\tau_c=1/R_c$ of OH* 
is thus about 0.4 s, much longer than the radiative lifetime 
of OH* expected to be about 10$^{-6}$ s 
\citep{BeckerHaaks1973}. 
Therefore collisional quenching does not significantly de-excite
the OH* radicals. 

\section{Summary}

    \setcitestyle{citesep={;}}
    
The EPOXI mission allowed us to investigate the innermost coma of 103P/Hartley 2, 
with high spatial resolution, revealing regions and processes usually 
unobservable with ground based observations. 

We analyzed the dust colors using the color filters mounted on 
the MRI camera and found that the dust coma in the first 5 km from the 
nucleus is bluer in the sunward
direction than in the tailward direction.
We attributed this to an enrichment in water ice in the sunward coma, as supported by 
HRI-IR spectral maps \citep{AHearn2011,Protopapa2014} and
the analysis of the large particles near the nucleus \citep{Kelley2013}. 
The bulk dust reddening is approximately 15\%/100 nm between 345 and 759 nm, 
in agreement with ground based observations
\citep{Knight2013,Lara2011}.

The analysis of 153 images acquired with the MRI-OH narrowband filters 
allowed us
to study the OH spatial distribution from very small projected distances 
to 10$^4$ km from the nucleus. OH has a clear
anti-sunward enhancement when observed at spatial scales larger than 1.2 km pix$^{-1}$
while it shows a much more isotropic distribution at smaller scales. 
The images of the innermost coma, acquired at CA with spatial scale about 80 m pix$^{-1}$,
show a significant radial sunward jet
extending up to 12 km from the nucleus and a second fountain-like structure to the right of the central plume. This structure resembles more a parent species 
distribution rather than a spatial distribution typically associated with fragment species. This feature is also very well correlated with the water vapor
distribution observed with HRI-IR \cite{AHearn2011}.

The OH column density radial profiles for the full dataset between 
10 and a few thousand km from the surface show a consistent behavior without significant discontinuity.
A two-generation Haser model is adequate to describe the observed profiles,
apart from the very innermost regions (inside 20 km), and in particular in
the sunward direction, which shows a steeper behavior than expected.
The derived water production rate curve, generally in good agreement with the
dust light curve, shows an excess flux at closest approach
that can not be explained with the sole increased resolution at CA.

The morphology and the flux excess are independent indications that OH fluorescence is not the 
sole emission process responsible for the OH brightness observed. 
Instead, prompt emission 
from excited OH*$(A^2\it{\Sigma}^+)$ molecules produced directly by the photodissociation
of water is likely responsible for the OH inner coma structure observed 
in MRI closest approach images. Using an appended Haser model, we 
were able to explain the observed flux excess within a factor 2, which is very reasonable 
considering the large uncertainties in the parameters assumed and 
the physical processes not considered in the calculations.
If this scenario applies, this would be the first time
that OH* NUV prompt emission is directly imaged through a narrowband filter
and its distribution is a direct tracer for water distribution in the coma 
of Hartley 2 at larger distances than what reachable with HRI-IR spectral maps
at CA.

\subsection{Future applications}

 Given that prompt emission of OH is directly related to the 
distribution of water, this allows for imaging the distribution
of water in optical wavelenghts, 
commonly accessible by CCDs on telescopes on Earth and flown on telescopes.

\begin{deluxetable}{llccc}
\tablenum{1}
\tablecaption{Optimization of an OH-PE dedicated filter. \label{tab:pefilter}}
\tablewidth{0pt}
\tablehead{
\colhead{} & \colhead{$\lambda_c$} & \colhead{$\Delta\lambda$} & \colhead{RFE} & \colhead{PE}\\
\colhead{} & \colhead{nm} & \colhead{nm} & \colhead{\%} & \colhead{\%}
}
\startdata
MRI-filter & 309.5 & 6.2 & 83.5 & 39.0\\
	   & 310.7 & 6.2 & 82.5 & 41.4\\
	   & 311.7 & 6.2 & 74.0 & 42.3\\
	   & 312.7 & 6.2 & 52.9 & 42.9\\
	   & 313.7 & 6.2 & 31.6 & 43.9\\
	   & 314.7 & 6.2 & 21.3 & 45.3\\
	   & 315.7 & 6.2 & 17.0 & 44.6\\
	   & 316.7 & 6.2 & 14.6 & 43.5\\
	   & 317.7 & 6.2 & 11.8 & 42.4\\
	   & 318.7 & 6.2 & 9.2 & 39.8\\
	   & 318.7 & 6.8 & 8.5 & 40.6\\
	   & 318.7 & 7.4 & 9.4 & 41.7\\
	   & 318.7 & 8.1 & 10.7 & 43.0\\
	   & 318.8 & 8.7 & 12.1 & 46.3\\
\enddata

 \end{deluxetable}

 \begin{figure}[hb!]
\epsscale{1.2}
\plotone{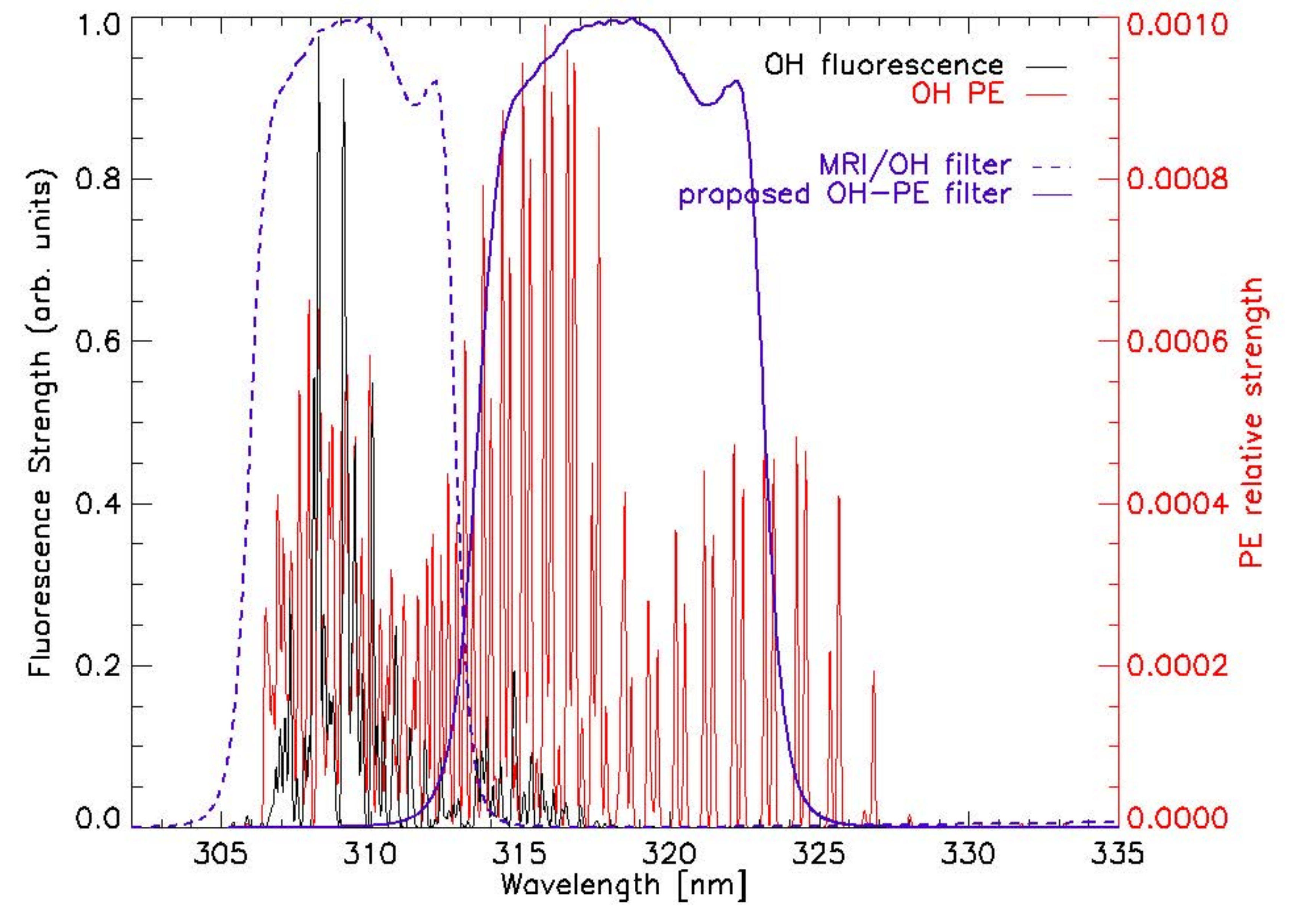}
    \caption{\small Bandpass of the proposed OH-PE dedicated
    filter (solid line) shifted and scaled with respect to 
    MRI-OH filter (dashed line) to have $\lambda_c=$318.8 nm
    and $\Delta\lambda=$8.7 nm to maximize prompt emission (red spectrum)
    contribute and minimize flurescence emission (black spectrum) contribute.}
    \label{fig:PEdedicated}
\end{figure}

\begin{figure}[ht!]
\epsscale{1.2}
\plotone{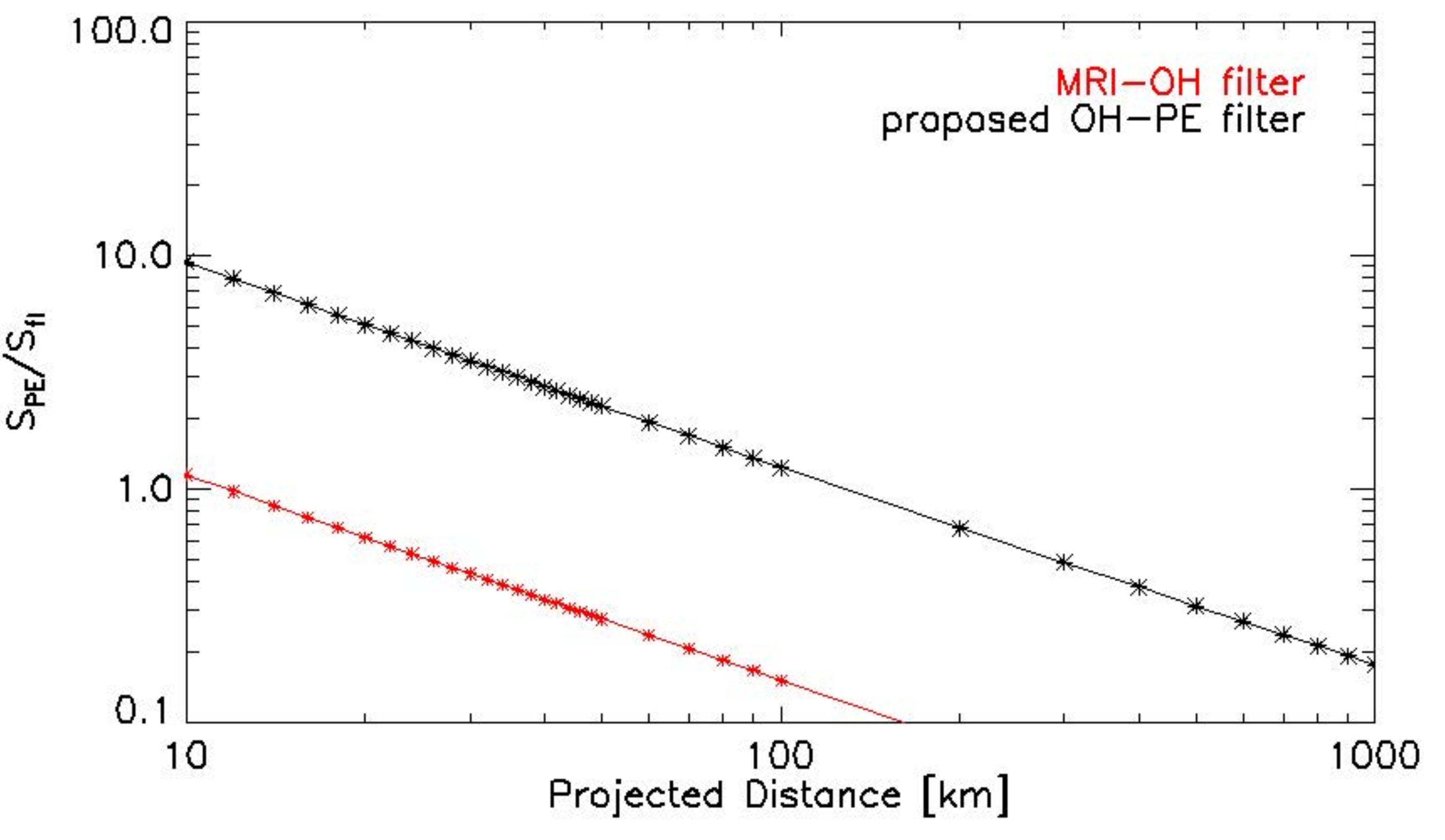}
    \caption{\small Relative strenght of OH-PE to 
    RFE emission as function of the projected distance 
    from the nucleus as seen through MRI-OH filter (red) and 
    the proposed OH-PE dedicated filter.}
    \label{fig:PEdedicated_relstrenght}
\end{figure}

 We demonstrated that a dedicated OH-PE filter
could be used to 
directly image the distribution of H$_2$O in the inner coma 
when the first 200 km of the inner coma can be resolved.
OH* prompt emission traces water better than forbidden [OI] emission 
which is more difficult to interpret 
\citep[see for example][]{Bodewits2016} because it can have multiple parents 
(H$_2$O, CO$_2$, CO, and O$_2$), and because the long lifetime 
of the $^1$D state leads to transport and quenching. 

We tentatively performed an optimization of the design 
of such a filter using MRI-OH 
filter profile shifted and scaled to minimize the contribution 
of fluorescence while maximizing prompt emission.
In Table \ref{tab:pefilter} we listed the percentage of total fluorescence 
(RFE) and the percentage of total prompt emission (PE) passing through 
the filter for each combination of central wavelength ($\lambda_c$) and 
bandwidth ($\Delta\lambda$). The first line shows the values for 
MRI-OH filter.
The two bands are partially superimposed, therefore a complete 
exclusion of fluorescence is difficult unless a sharper filter cutoff is used at low wavelenghts.
The best solution for a filter profile similar to MRI-OH 
is given by a filter centered at 318.8 nm with 
bandwidth of 8.7 nm (Fig. \ref{fig:PEdedicated}) which would
include 12.1\% of total fluorescence emission 
and 46.3\% of total prompt emission.
This is considered the best compromise to include the PE peak and 
exclude as much fluorescence as possible.

We computed the relative strenght of the two bands for the conditions
at Hartley 2 through such a filer (Fig. \ref{fig:PEdedicated_relstrenght}).
For comparison the relative strenght through MRI-OH filter is also shown.
Fig. \ref{fig:PEdedicated_relstrenght} revelas that through such 
OH-PE dedicated filter the prompt emission would be ten times 
stronger than fluorescence in the inner coma, at about 10 km from the nucleus.
This result confirms the importance of including a dedicated 
OH-PE filter in the payload of future missions to comets to allow a direct
measurement of the distribution of water at optical waveleghts.

\acknowledgments
\noindent {\bf Acknowledgments}

\noindent We would like to dedicate this work to Michael F. A'Hearn, 
who deceased on 
May 29, 2017 a few days before the submission of this manuscript,
and whose insight, expertise and mentoring has been crucial for the conception
and develompment of this research.
DB, TLF, and MSPK were supported by PMDAP grant 11-PMDAP11-0039.

 \bibliographystyle{aasjournal}
 \bibliography{biblio5}

\end{document}